\documentclass[
reprint,
amsmath,
amssymb,
aps,
prl
]{revtex4-2} 

\usepackage{graphicx}
\usepackage{upgreek}
\usepackage{dcolumn}
\usepackage{bm}
\usepackage{siunitx}
\usepackage{braket}
\usepackage[dvipsnames]{xcolor}
\usepackage[normalem]{ulem}
\usepackage[utf8]{inputenc}
\usepackage{xargs}
\usepackage[colorinlistoftodos,prependcaption,textsize=tiny]{todonotes}

\usepackage{hyperref}
\hypersetup{
    colorlinks,
    linkcolor={black},
    citecolor={blue!80!black},
    urlcolor={blue!80!black}
}

\begin{document}

\preprint{APS/123-QED}

\title{Emergence of second-order coherence in superfluorescence}

\author{Constanze Bach}
\author{Felix Tebbenjohanns}
\author{Christian Liedl}
\author{Philipp Schneeweiss}
\author{Arno Rauschenbeutel}

\affiliation{Department of Physics, Humboldt-Universität zu Berlin, 10099 Berlin, Germany}

\date{\today}

\begin{abstract}

We experimentally investigate the second-order quantum coherence function of a superradiant burst in a cascaded quantum system. We chirally (i.e.~direction-dependently) couple roughly 900 cesium atoms to the forward propagating mode of an optical nanofiber. 
We then prepare the ensemble in the maximally inverted state, where the subsequent collective emission of a burst is known as superfluorescence.
Here, we observe that second-order coherence emerges in the course of the decay.
This is a clear feature of the underlying collective dynamics that is also at the origin of the superradiant burst itself. We furthermore study the dynamics of the second-order coherence function of the emission in dependence on the initial average dipole moment of the ensemble.
In addition, by correlating the detection of early and late photon emission events, we obtain evidence for fundamental shot-to-shot fluctuations in the delay of the start of the burst emission.
Our findings reveal that, despite the fundamentally different coupling Hamiltonian, superradiance in cascaded and symmetrically coupled systems feature a strikingly large number of similarities.
\end{abstract}

\maketitle

\paragraph{Introduction.}
The collective emission of radiation is an ubiquitous physical process that underlies devices such as lasers and phased-array antennas, applications such as optical quantum memories, and even phenomena in space such as astrophysical cyclotron masers~\cite{letokhov2009astrophysical,guerin2017light,sheremet2023waveguide}. Already in 1954, R.~H.~Dicke studied the collective emission originating from a dense ensemble of two-level quantum emitters that are initially prepared in the maximally excited state~\cite{dicke1954coherence, skribanowitz1973observation, gross1976observation, gross1982superradiance}. In this system, superfluorescence occurs, where the radiated optical power first increases with time, then reaches a maximum and, eventually, decays. 
Such a superradiant burst of light is a hallmark effect in many-body quantum optics and qualitatively different from the exponential decay observed in the emission from independent atoms. Only recently, it was theoretically~\cite{cardenaslopez2023manybody} and experimentally~\cite{liedl2024observation} shown that a burst also occurs for a cascaded interaction between emitters arranged in a chain. In this setting, a given atom $i$ influences the decay of atoms $j>i$ but not vice versa~\cite{gardiner1993driving, carmichael1993quantum}. 
Cascaded interactions between quantum emitters can be engineered, for example with Rydberg atoms~\cite{stiesdal2021controlled}, acousto-optic control techniques~\cite{calajo2019quantum}, spin-orbit coupled Bose-Einstein condensates~\cite{ramos2014quantum}, and occur naturally for quantum emitters in optical near fields due to chiral light-matter interaction~\cite{lodahl2017chiral}. 
While superfluorescence and in general superradiant phenomena have been studied extensively~\cite{gross1982superradiance, bohnet2011steadystate,
bienaime2013cooperativity,cong2016dicke,
laske2019pulse,
ferioli2021laser,pennetta2022observation,reitz2022cooperative, santos2023generation, ferioli2024non, ostermann2024breakdown}, important properties such as the coherence and the photon statistics of the burst emission have only been explored sparsely in the literature~\cite{haake1972quantum,lopes2014secondorder,jahnke2016giant,stryzhenko2024Nscaling,liedl2024observation}. 

Here, we experimentally study the two-time second-order quantum coherence function~\cite{glauber1963quantum}, $g^{(2)}(t_1,t_2)$, of the superradiant burst emitted by a cascaded quantum system. Our system is realized with 900 cesium atoms chirally coupled to the guided mode of an optical nanofiber, i.e., a cylindrical dielectric waveguide with a diameter smaller than the wavelength of the guided light. 
From these second-order correlations, we discern the regimes of superfluorescence and superradiance, we directly observe the spontaneous build-up of second-order coherence of the radiation emitted by an ensemble of initially independent atoms, and we infer shot-to-shot fluctuations of the delay between the excitation and the burst emission. 
We compare our data to the symmetric Dicke model as well as to a stochastic simulation of our system, which is based on the truncated Wigner approximation.

\begin{figure}
    \centering
    \includegraphics[width=0.9\columnwidth]{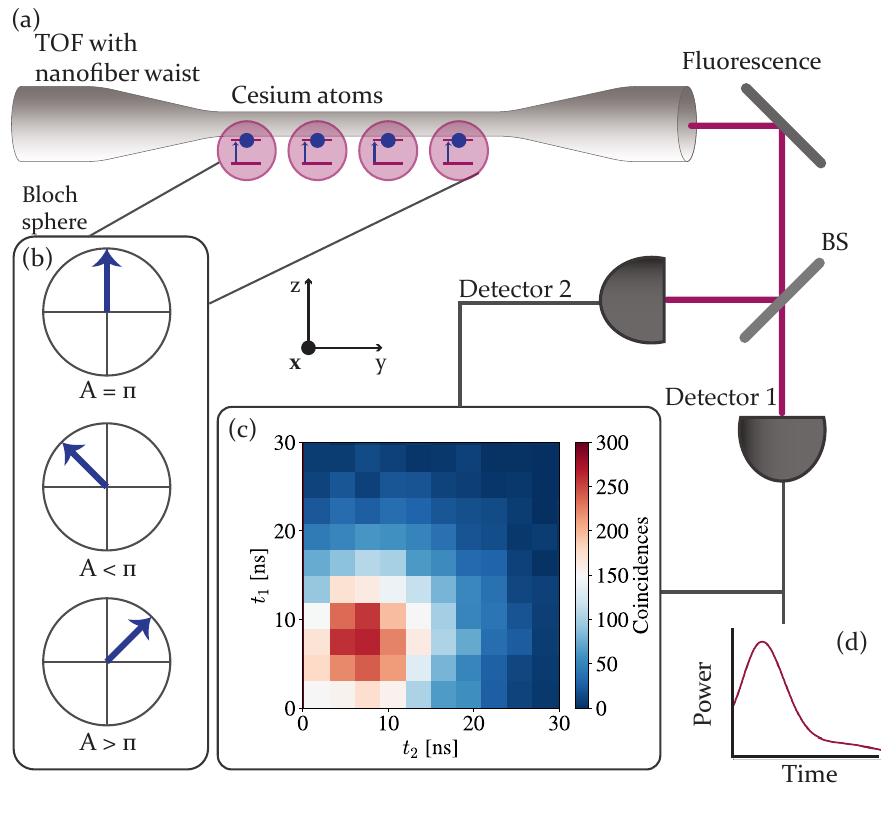}
    \caption{(a) Schematic of the experimental setup for measuring the second-order coherence function $g^{(2)}(t_1,t_2)$ of a superradiant burst. An ensemble of about 900 cesium atoms (red circles) are optically trapped near the surface of an optical nanofiber.
    A pulse of nanofiber-guided resonant light coherently excites the ensemble close to full inversion of a cycling transition which predominantly re-emits the light into the forward-propagating nanofiber mode, thereby realizing a cascaded quantum system. The optical power and its second-order correlation are obtained using a Hanbury-Brown and Twiss-type detection setup. (b) Representation of the initial state, parameterized by the Rabi pulse area $A$ on the Bloch sphere. (c) Measured two-photon coincidence rates within a $3$~ns bin width. A maximum of roughly $250$ two-photon coincidences occurs at $t_1=t_2\approx 7.5~\mathrm{ns}$ for a total of $10^7$ excitation pulses. (d) Sketch of the optical power on a single detector. }
    \label{fig:setup}
\end{figure}

\paragraph{Setup and Measurement.}
Our experimental setup is sketched in Fig.~\ref{fig:setup}.
We realize an optical trapping potential in the evanescent field surrounding the nanofiber-waist of a tapered optical fiber (TOF), by sending running wave blue-detuned laser light (wavelength 760 nm, power $20.5~\mathrm{mW}$) and forward and backward propagating red-detuned laser light (wavelength 1064 nm, powers $ 1.3~\mathrm{mW}$ and $1.1~\mathrm{mW}$, respectively) through the nanofiber. 
This creates two diametral arrays of trapping sites~\cite{vetsch2010optical}, which we probabilistically load with cold cesium atoms by overlapping the nanofiber with a molasses-cooled cloud of atoms from a magneto-optical trap (MOT). 
Due to the collisional blockade effect, each trapping site is filled with at most a single atom~\cite{Schlosser2002}. 
We prepare the atoms on only one side of the fiber in their motional ground state using side-selective degenerate Raman cooling (DRC) for $50$~ms~\cite{Meng2018}.
With this, we end up with a single one-dimensional array of trapped cesium atoms, which lies in the $x-y$-plane, see Fig.~\ref{fig:setup}(a). Upon DRC, the internal state of the atoms is $\ket{g} =\ket{6S_{1/2}, \mathrm{F} = 4, m_\mathrm{F} = -4}$, where the quantization axis is chosen along the z-direction. 
We control the number of atoms, $N$, via the MOT loading time and determine it by measuring the optical depth (OD) with transmission spectroscopy through the TOF. 
All the measurements shown here are performed at an OD of about $40$, corresponding to $N \approx 900$~\cite{vetsch2010optical}. 

Because of spin-momentum locking~\cite{mitsch2014quantum}, the polarization of the waveguide mode at the position of the atoms depends strongly on its sense of propagation, forward or backward.
In particular, the radiation emitted on the $\sigma^-$-polarized atomic D2-cycling transition $\ket{g}\rightarrow\ket{e} = \ket{6P_{3/2}, \mathrm{F} = 5, m_\mathrm{F} = -5}$ has an overlap with the forward-propagating mode of $\beta\approx 0.01$, while its overlap with the backward-propagating mode is about ten times smaller.
Thanks to this so-called chiral coupling, the nanofiber-coupled atoms thus realize a cascaded quantum system~\cite{carmichael1993quantum, gardiner1993driving, lodahl2017chiral}.
We coherently drive the  $\ket{g} \rightarrow \ket{e}$ transition with a fiber-guided forward propagating resonant optical Rabi pulse of duration $T_\mathrm{pulse} = 4~\mathrm{ns}$, which is shorter than the excited-state lifetime of $\tau = 30.5~\mathrm{ns}$~\cite{liedl2023collective}.
This prepares the ensemble close to the product state
\begin{equation}
    \ket{\psi_0} = \bigotimes_{k=1}^N \left[\cos \left(  \frac{A}{2}\right) \ket {g_k} - i  \sin \left(  \frac{A}{2}\right) \ket {e_k}\right] ,
    \label{eq:product_state}
\end{equation}
where the atomic index $k=1,\dots,N$ increases along the propagation direction of the light.
This initial state is depicted on the Bloch sphere in Fig.~\ref{fig:setup}(b).
The Rabi pulse area $A=\Omega T_\mathrm{pulse}$ is defined by the Rabi frequency $\Omega$ applied to the first atom. 
In practice, because of absorption of the pulse upon propagation through the atomic ensemble and due to inhomogeneous atom-waveguide coupling, the atoms experience slightly different pulse areas $A_k\approx A$, see discussions in Refs.~\cite{liedl2023collective, liedl2024observation}.

In each experimental run, we repeatedly excite the ensemble 400 times within $80~\mathrm{ms}$.
During this probing, which would slightly heat the atoms, we continuously cool the ensemble by applying DRC on the D1 line ($\ket{6S_{1/2}, \mathrm{F} = 4} \rightarrow \ket{6P_{1/2}, \mathrm{F} = 4}$) with a free-space laser, achieving a survival of $75\%$ of the atoms by the end of the probing period.
Because the scattering rate of the D1 light is sufficiently small, this cooling does not alter the burst dynamics.
A fully inverted and sufficiently large ensemble (for us $N \gg 100$ and $A\approx \pi$) then radiates a superradiant burst as sketched in Fig. 1(d) and experimentally investigated in detail in Ref.~\cite{liedl2024observation}. 
A hybrid photodetector (Hamamatsu, R10467, dead time $<2~\mathrm{ns}$) captures both this superradiant burst and the preceding transmitted excitation Rabi pulse. 
For the two-time correlation measurement of the burst, we split the light into two parts and delay one half by about $100~\mathrm{ns}$ using a $20~\mathrm{m}$ long optical fiber before both fractions reach the same detector. 
Since the decay dynamics is much faster than $100~\mathrm{ns}$, this allows us to measure the two-photon coincidence rate, $n_c(t_1,t_2)$.
Here, $t_1$ and $t_2$ are measured from the end of the Rabi pulse.
A sample measurement of $n_c(t_1, t_2)$ during the emission of a superradiant burst for a maximally inverted ensemble is shown in Fig.~\ref{fig:setup}(c), with a binning of $3~\mathrm{ns}$. 
The normalized second-order coherence function is then given by 
 \begin{equation}
    g^{(2)}(t_1, t_2) = \frac{n_c(t_1,t_2)}{n_1(t_1)n_2(t_2)},
    \label{eq:g2_rates}
\end{equation}
where $n_{j}(t_{j})$ is the photon rate on detector $j$ at the respective time $t_{j}$ with $j = 1,2$. 

\begin{figure}
    \centering
    \includegraphics[width=0.85\columnwidth, trim={0.5cm 0.7cm 0 0}]{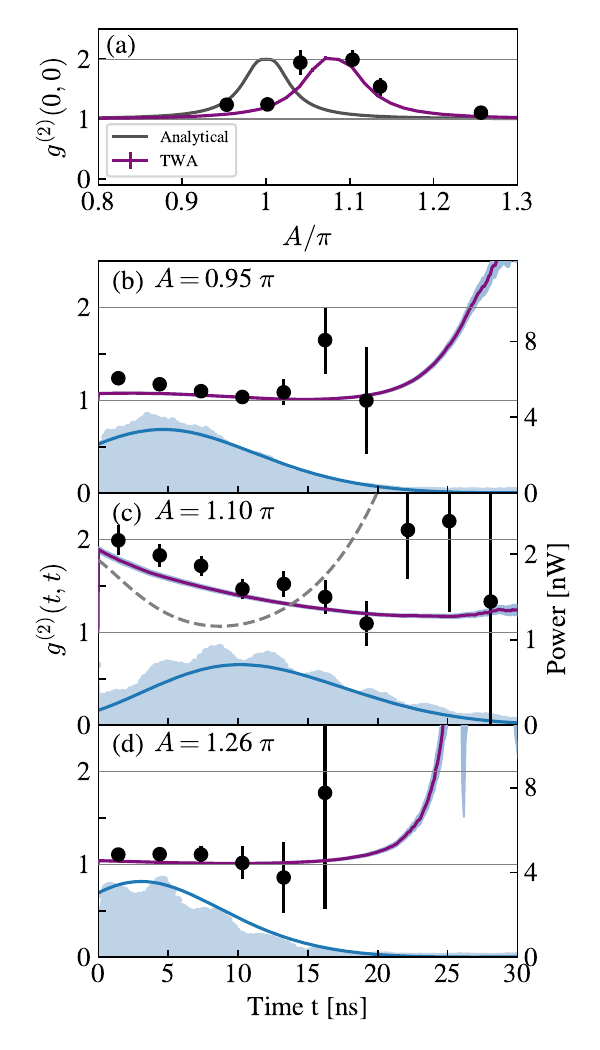}
    \caption{
    (a) Experimentally measured second-order coherence function $g^{(2)}(0, 0)$ of the initial emission as a function of the excitation Rabi pulse area $A$ (back circles with Poissonian error bars). 
    The grey and purple theory lines are, respectively, an analytical prediction according to Eq.~\eqref{eq:g2_pr} and a numerical prediction of the TWA model for $N=900$ atoms. 
    (b)-(d) $g^{(2)}(t, t)$ as a function of time for three values of $A$, as indicated. For $A = 1.10\pi$, $g^{(2)}(t, t)$ develops from 2 to 1 during the burst. For $A \neq 1.10\pi$, $g^{(2)}(t, t)=1$ is constant. The purple lines show TWA predictions with the one-sigma error due to a finite number of computed trajectories (shaded areas). The grey dashed line in panel (c) is the prediction by the symmetric Dicke model with $N_\text{Dicke}=9$.
    The optical power $P(t)$ is shown as the light blue area with the corresponding TWA prediction as the dark blue line. The oscillations on top of $P(t)$ are quantum beats that originate from weak excitation of the hyperfine state $\ket{6P_{3/2}, \mathrm{F} = 4}$.
    In (b) - (d) we omitted datapoints, where we did not observe a single coincidence.
    }
    \label{fig:g2_vs_A}
\end{figure}

\paragraph{Initial second-order coherence function.}
Let us first discuss the value of the second-order coherence function of the light that is emitted right after the excitation pulse, $g^{(2)}(0,0)$, which is shown in Fig.~\ref{fig:g2_vs_A}(a) as a function of $A$ (black circles). 
The data reaches a peak value of $g^{(2)}(0,0)=2$ for $A \approx 1.1\pi$, falls off over a range of about $\pm 0.1\pi$, and decreases to $g^{(2)}(0,0)\approx1$ for substantially smaller or larger pulse areas. 
This behavior can be understood in the following way. As discussed in Ref.~\cite{liedl2023collective}, because of the absorption of the pulse along the ensemble, a pulse area at the first atom slightly larger than $\pi$ (here, $A \approx 1.07\pi$) maximizes the mean excitation stored in the ensemble. 
For this case of maximum inversion, i.e. $\ket{\psi_0}\approx \bigotimes_k\ket{e_k}$, the ensemble-averaged dipole moment vanishes, see also Bloch sphere representation in Fig.~\ref{fig:setup}(b).
Correspondingly, the fields radiated by the atoms do not have a fixed phase relationship. 
Thus, the initially emitted light features the statistics of independent atoms, yielding $g^{(2)}(0,0) = 2$.
By contrast, when exciting the ensemble with a pulse area $A$ below or above maximal inversion, a non-zero average dipole moment is imprinted on the ensemble by the excitation laser.
Thus, the atoms radiate in phase and the statistics of the emitted light reproduces those of the excitation laser field, resulting in $g^{(2)}(0,0)=1$.

Let us now theoretically analyze $g^{(2)}(0,0)$ when the atoms are prepared in $\ket{\psi_0}$ as defined in Eq.~\eqref{eq:product_state}.
The total electric field $\hat{E}$ in the detected mode is given by the sum over the contribution of all atomic dipoles $\hat{\sigma}_k=\ket{g_k}\bra{e_k}$, i.e., $\hat E\propto \sum_k \hat \sigma_k$.
As shown in the supplemental material (SM), we find the following expression for second-order quantum coherence function
\begin{equation}
    g^{(2)}(0,0) = \frac{\bra{\psi_0}\hat E^\dagger\hat E^\dagger\hat E\hat E \ket{\psi_0}}{\bra{\psi_0}\hat E^\dagger \hat E  \ket{\psi_0}^2} \approx 2 - \frac{1}{(1+\frac{1}{N\cos^2(A/2)})^2},
    \label{eq:g2_pr}
\end{equation}
where the right hand side is exact in the large-ensemble limit of
$N\gg1$. We show this prediction as a grey line for $N=900$ in Fig.~\ref{fig:g2_vs_A}(a).
The width of the peak predicted for $g^{(2)}(0,0)$ as a function of $A$ is approximately $2\pi / \sqrt{N}$, which roughly matches our data. 
We can additionally model the absorption of the Rabi pulse along the ensemble by applying a stochastic numerical model based on the truncated Wigner approximation (TWA)~\cite{mink2023collective, tebbenjohanns2024predicting}, see purple line in Fig.~\ref{fig:g2_vs_A}(a). It is apparent that this explains the shift of the peak by $0.07\pi$. More details on the TWA follow below.

\paragraph{Dynamics of second-order coherence function.}
In Figs.~\ref{fig:g2_vs_A}(b)--(d), we show the measured optical power, which is averaged over more than $10^6$ excitation pulses, as the blue shaded area, and the normalized second-order coherence function at equal time, $g^{(2)}(t,t)$, during the burst as black circles for different initial atomic states characterized by the Rabi pulse area $A$.
The characteristic feature of the superradiant decay of a maximally inverted atomic ensemble is the burst of the optical power~\cite{liedl2024observation}. It is most prominent in panel (c), where the initial state is close to maximal inversion.
Notably, there we measure a decrease of $g^{(2)}(t,t)$ during the burst emission from its initial value of $g^{(2)}(0,0)=2$ to $g^{(2)}(t,t)=1$ at $t\approx 19~\mathrm{ns}$.
In stark contrast, we observe a constant $g^{(2)}(t,t)\approx 1$ for the entire duration of the burst emission in panels (b) and (d), where the ensemble's average initial dipole moment is nonzero.

The theoretical modelling of such second-order correlations for as many as one thousand atoms requires state-of-the-art approaches~\cite{ostermann2012collective,caneva2015quantumdynamics,
mahmoodian2020dynamics, 
robicheaux2021beyond,
arranzregidor2021modeling,
kusmierek2023higherorder, 
cardenaslopez2023manybody,
rubiesbigorda2023dynamic,
kleinbeck2023creation,
vlasiuk2023two}.
Here, we apply an efficient, recently developed stochastic simulation tool, which is based on the truncated Wigner approximation for spins (TWA)~\cite{schachenmayer2015manybody, mink2023collective}. The latter can describe many emitters and many excitations. In our model, the system dynamics are mapped to a set of $2N$ coupled, nonlinear stochastic differential equations, which can be numerically solved for thousands of atoms, to accurately predict both the output power (blue lines in Fig.~\ref{fig:g2_vs_A}) and $g^{(2)}(t,t)$ (purple lines).
For the details on the implementation of the TWA, see Ref.~\cite{tebbenjohanns2024predicting}.
Our modelling includes the coherent Rabi excitation process, where we account for the absorption of the excitation pulse along the ensemble as well as for the inhomogeneous atom-waveguide coupling due to the atom's thermal motion.
In addition to predicting the system dynamics with the TWA model, we also numerically solve the symmetric Dicke model and present its prediction for $g^{(2)}(t,t)$ as the grey dashed line in Fig.~\ref{fig:g2_vs_A}(c).
Since this model assumes perfect atom-waveguide coupling ($\beta_\text{Dicke}=1$), we use an atom number of $N_\text{Dicke} = 9 \approx \beta N$~\cite{ferioli2024non} (see SM).
This model shows qualitative agreement with our experimental data, particularly regarding the reduction of $g^{(2)}(t,t)$ from $2(1-1/N_\text{Dicke}) \approx 2$ to $1$. However, it does not capture the experimentally observed time dynamics as accurately as the TWA.

For maximal inversion, the ensemble has no dipole moment. Therefore the emission is seeded by vacuum fluctuations, a situation in which superradiance is also termed superfluorescence~\cite{vrehen1980superfluorescence, bonifacio1975cooperative}. 
Following the initially independent emission, second-order coherence builds up while the light is radiated.
This characteristic dynamics of superfluorescence occurs because the atomic dipoles spontaneously synchronize via the shared mode.
More in detail, we take the fact that we observe $g^{(2)}(t,t)<2$ as a signature of this synchronization, because it violates the Siegert relation~\cite{ferreira2020connecting}, which holds for independent emitters and reads $g^{(2)}(t,t)=1+|g^{(1)}(t,t)|^2 =2$~\footnote{The normalized, equal-time first-order coherence function is by definition equal to unity, $g^{(1)}(t,t)=1$~\cite{glauber1963quantum, loudon2000quantum}}.
Note that for an excitation Rabi pulse area sufficiently below and above maximal inversion [c.f. Fig.~\ref{fig:g2_vs_A}(a) and (c)], the Siegert relation is also broken. However, there the synchronization of the dipoles does not spontaneously build up, but is already present at $t=0$ since the atomic dipoles are already synchronized to the excitation laser~\cite{liedl2024observation}.

\begin{figure}
    \centering
    \includegraphics[width=0.9\columnwidth, trim={0.8cm 0.5cm 0 0}]{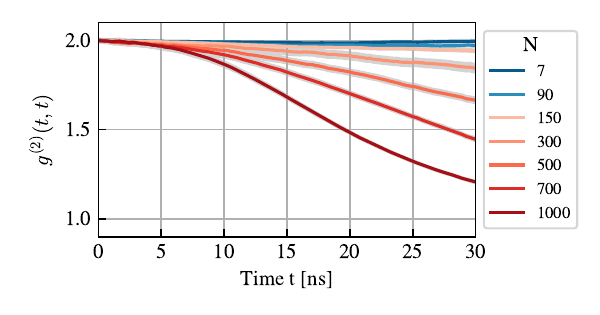}
    \caption{
    TWA simulation of $g^{(2)}(t,t)$ for different numbers of atoms. For $N<N_\mathrm{thr}\approx 100$ (blue) no burst occurs~\cite{liedl2024observation} and $g^{(2)}(t,t) = 2$ is constant. For $N>N_\mathrm{thr}$ (red), $g^{(2)}(t,t)$ evolves towards 1, i.e., second-order coherence builds up during the collective emission.}
    \label{fig:coherence}
\end{figure}

To further our understanding that the emergence of second-order coherence is a collective phenomenon, we now study $g^{(2)}(t,t)$ as a function of the number of atoms $N$.
In a cascaded system with imperfect atom-mode coupling ($\beta < 1$), a burst only appears when $N$ exceeds a threshold value, $N_{\text{thr}} = 1 + 1/\beta$, whereas the atoms decay independently for $N \ll N_{\text{thr}}$~\cite{gross1982superradiance,cardenaslopez2023manybody}. Here, $N_{\text{thr}}$ is about 100.
Due to low count rates, we cannot reliably measure $g^{(2)}(t,t)$ for small atom numbers in our experiment and therefore analyze this regime theoretically using the TWA model.
In Fig.~\ref{fig:coherence}, we show the calculated $g^{(2)}(t,t)$ for an ensemble of homogeneously coupled atoms with $\beta=0.01$, which is initialized in $\ket{\psi_0}= \bigotimes_k\ket{e_k}$.
The rate at which the second-order coherence builds up decreases when $N$ is reduced and, importantly, for $N<N_{\text{thr}}$ (depicted in blue), $g^{(2)}(t,t)$ is a constant. 
In that regime, the ensemble remains in a product state $ \hat\rho(t) = \bigotimes_k \hat\rho_k(t)$, where $\hat\rho_k(t)$ is the density matrix of the $k^\text{th}$ atom, yielding $ g^{(2)}(t,t) = g^{(2)}(0,0)=2 $. 
These simulation results further support the conclusion that a build-up of second-order coherence in our experiment is a signature of a departure of the ensemble from the product state, i.e. $ \hat\rho(t) \neq  \bigotimes_k \hat\rho_k(t)$.

\begin{figure}
    \centering
    \includegraphics[width=0.9\columnwidth]{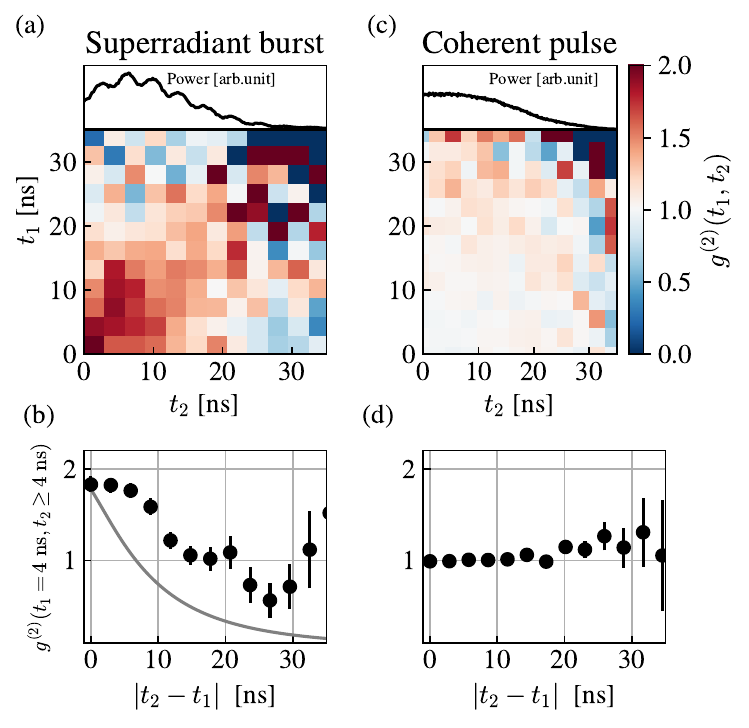}
    \caption{(a) Color plot of $g^{(2)}(t_1, t_2)$ of the superradiant burst at detection times $t_1, t_2$. Red (blue) colors indicate $g^{(2)}>1$ ($g^{(2)}<1$). 
    The trace of the average measured power (black line) is shown in the top panel for comparison.
    (b) A cut of $g^{(2)}(t_1=4~\mathrm{ns}, t_2)$ as a function of $|t_2 - t_1|$ (black circles with Poissonian error bars), reveals anti-correlations for delays above $20~\mathrm{ns}$. The grey solid line is a prediction based on the symmetric Dicke model for $N = 9$ and $\beta = 1$. 
    (c), (d) Same analysis for a coherent laser pulse with similar shape and duration as the burst in (a). Here, $g^{(2)}(t_1, t_2)=1$ within the error bars as the photons are uncorrelated.}
    \label{fig:bunches_v2}
\end{figure}

\paragraph{Shot-to-shot variations.}
Finally, we turn to the statistical properties of the light emitted from a maximally inverted ensemble by considering $g^{(2)}(t_1,t_2)$ for unequal times $t_1 \neq t_2$, depicted as a color plot in Fig.~\ref{fig:bunches_v2}(a). 
In Fig.~\ref{fig:bunches_v2}(b), we show $g^{(2)}(t_1=4~\mathrm{ns},t_2)$ as a function of the time difference $|t_2-t_1|$. This data thus quantifies the correlation between the burst light at early and late times.
Interestingly, the detection events are correlated,  $g^{(2)} (t_1,t_2) >1$, for small time differences and anti-correlated, $g^{(2)} (t_1,t_2) <1$, for larger time differences.
These anti-correlations are also apparent as blue regions in the bottom right and top left corner of Fig.~\ref{fig:bunches_v2}(a). 
For comparison, we present the results of the same correlation measurement performed with a coherent laser pulse of similar shape and duration in panels (c) and (d).
Here, as expected, $g^{(2)}(t_1, t_2)=1$ is constant within the error bars for all combinations of $t_1$ and $t_2$ as there are no correlations between the detection of early and late photons. 
For the time being, the TWA cannot predict two-time correlators.
In order to still compare our observations to a model prediction, we therefore apply the quantum regression theorem to the symmetric Dicke model for $N_\text{Dicke} = 9$ (see SM). The resulting prediction for $g^{(2)}(t_1=4~\mathrm{ns},t_2)$ is shown as a grey solid line in panel (b) and features a qualitatively similar transition from $g^{(2)}(t_1,t_1)\approx 2$ to $g^{(2)}(t_1=4~\mathrm{ns},t_2) < 1$ at a large time difference.

To understand these anti-correlations, let us assume that individual bursts have a duration $\tau_s$. Moreover, let us assume that, despite the identical preparation of the atoms, each burst occurs after a varying delay with respect to the excitation pulse~\cite{gross1982superradiance}. 
When averaging over many realizations, the resulting power trace will have a duration that is longer than $\tau_s$. At the same time, the probability of the detection of two photons separated by more that $\tau_s$ is smaller in a single shot than on average. We thus interpret the observed anti-correlations of $g^{(2)}(t_1,t_2)$ as a signature of these fluctuations. 
For the case of symmetric coupling, such shot-to-shot variations of the burst are a known feature of superfluorescence and originate from the fact that the burst is triggered by spontaneous emission~\cite{cong2016dicke, florian1984time}.

\paragraph{Conclusion and Outlook.}
In conclusion, we have measured and theoretically investigated the two-time second-order quantum coherence function, $g^{(2)}(t_1, t_2)$, of a superradiant burst emitted by a cascaded quantum system. 
This allowed us to observe the emergence of second-order coherence in the light emitted by initially independent emitters.
Remarkably, this occurs despite the absence of any feedback in a cascaded system.
In particular, in conjunction with the results from Ref.~\cite{liedl2024observation}, it becomes apparent that superradiance in a cascaded quantum system features surprisingly many similarities with the symmetric Dicke model.

Further characterization of the photonic output state of cascaded quantum systems includes measurements of even higher-order correlations, such as $g^{(3)}(t_1,t_2,t_3)$.
Moreover, we would like to investigate the emission of coherently and incoherently driven cascaded quantum systems in the steady-state.
From a theoretical standpoint, such experiments allow one to benchmark state-of-the-art simulation methods of many-body quantum systems. From a conceptional point of view, they may shed light on, e.g., the physics underlying superradiant lasing~\cite{bohnet2011steadystate}.

\begin{acknowledgments}
We thank M.~Fleischhauer, C.~Mink, K.~M{\o}lmer, A.~Poddubny, J.~Volz, L.~Yatsenko for fruitful discussions.
We acknowledge funding by the Alexander von Humboldt Foundation in the framework of the Alexander von Humboldt Professorship endowed by the Federal Ministry of Education and Research.
\end{acknowledgments}

\bibliography{bib}

\begin{thebibliography}{53}%
\makeatletter
\providecommand \@ifxundefined [1]{%
 \@ifx{#1\undefined}
}%
\providecommand \@ifnum [1]{%
 \ifnum #1\expandafter \@firstoftwo
 \else \expandafter \@secondoftwo
 \fi
}%
\providecommand \@ifx [1]{%
 \ifx #1\expandafter \@firstoftwo
 \else \expandafter \@secondoftwo
 \fi
}%
\providecommand \natexlab [1]{#1}%
\providecommand \enquote  [1]{``#1''}%
\providecommand \bibnamefont  [1]{#1}%
\providecommand \bibfnamefont [1]{#1}%
\providecommand \citenamefont [1]{#1}%
\providecommand \href@noop [0]{\@secondoftwo}%
\providecommand \href [0]{\begingroup \@sanitize@url \@href}%
\providecommand \@href[1]{\@@startlink{#1}\@@href}%
\providecommand \@@href[1]{\endgroup#1\@@endlink}%
\providecommand \@sanitize@url [0]{\catcode `\\12\catcode `\$12\catcode
  `\&12\catcode `\#12\catcode `\^12\catcode `\_12\catcode `\%12\relax}%
\providecommand \@@startlink[1]{}%
\providecommand \@@endlink[0]{}%
\providecommand \url  [0]{\begingroup\@sanitize@url \@url }%
\providecommand \@url [1]{\endgroup\@href {#1}{\urlprefix }}%
\providecommand \urlprefix  [0]{URL }%
\providecommand \Eprint [0]{\href }%
\providecommand \doibase [0]{https://doi.org/}%
\providecommand \selectlanguage [0]{\@gobble}%
\providecommand \bibinfo  [0]{\@secondoftwo}%
\providecommand \bibfield  [0]{\@secondoftwo}%
\providecommand \translation [1]{[#1]}%
\providecommand \BibitemOpen [0]{}%
\providecommand \bibitemStop [0]{}%
\providecommand \bibitemNoStop [0]{.\EOS\space}%
\providecommand \EOS [0]{\spacefactor3000\relax}%
\providecommand \BibitemShut  [1]{\csname bibitem#1\endcsname}%
\let\auto@bib@innerbib\@empty
\bibitem [{\citenamefont {Letokhov}\ and\ \citenamefont
  {Johansson}(2009)}]{letokhov2009astrophysical}%
  \BibitemOpen
  \bibfield  {author} {\bibinfo {author} {\bibfnamefont {V.}~\bibnamefont
  {Letokhov}}\ and\ \bibinfo {author} {\bibfnamefont {S.}~\bibnamefont
  {Johansson}},\ }\href@noop {} {\emph {\bibinfo {title} {Astrophysical
  lasers}}}\ (\bibinfo  {publisher} {Oxford University Press},\ \bibinfo {year}
  {2009})\BibitemShut {NoStop}%
\bibitem [{\citenamefont {W.~Guerin}\ and\ \citenamefont
  {Kaiser}(2017)}]{guerin2017light}%
  \BibitemOpen
  \bibfield  {author} {\bibinfo {author} {\bibfnamefont {M.~R.}\ \bibnamefont
  {W.~Guerin}}\ and\ \bibinfo {author} {\bibfnamefont {R.}~\bibnamefont
  {Kaiser}},\ }\bibfield  {title} {\bibinfo {title} {Light interacting with
  atomic ensembles: collective, cooperative and mesoscopic effects},\ }\href
  {https://doi.org/10.1080/09500340.2016.1215564} {\bibfield  {journal}
  {\bibinfo  {journal} {Journal of Modern Optics}\ }\textbf {\bibinfo {volume}
  {64}},\ \bibinfo {pages} {895} (\bibinfo {year} {2017})}\BibitemShut
  {NoStop}%
\bibitem [{\citenamefont {Sheremet}\ \emph {et~al.}(2023)\citenamefont
  {Sheremet}, \citenamefont {Petrov}, \citenamefont {Iorsh}, \citenamefont
  {Poshakinskiy},\ and\ \citenamefont {Poddubny}}]{sheremet2023waveguide}%
  \BibitemOpen
  \bibfield  {author} {\bibinfo {author} {\bibfnamefont {A.~S.}\ \bibnamefont
  {Sheremet}}, \bibinfo {author} {\bibfnamefont {M.~I.}\ \bibnamefont
  {Petrov}}, \bibinfo {author} {\bibfnamefont {I.~V.}\ \bibnamefont {Iorsh}},
  \bibinfo {author} {\bibfnamefont {A.~V.}\ \bibnamefont {Poshakinskiy}},\ and\
  \bibinfo {author} {\bibfnamefont {A.~N.}\ \bibnamefont {Poddubny}},\
  }\bibfield  {title} {\bibinfo {title} {Waveguide quantum electrodynamics:
  Collective radiance and photon-photon correlations},\ }\href
  {https://doi.org/10.1103/RevModPhys.95.015002} {\bibfield  {journal}
  {\bibinfo  {journal} {Rev. Mod. Phys.}\ }\textbf {\bibinfo {volume} {95}},\
  \bibinfo {pages} {015002} (\bibinfo {year} {2023})}\BibitemShut {NoStop}%
\bibitem [{\citenamefont {Dicke}(1954)}]{dicke1954coherence}%
  \BibitemOpen
  \bibfield  {author} {\bibinfo {author} {\bibfnamefont {R.~H.}\ \bibnamefont
  {Dicke}},\ }\bibfield  {title} {\bibinfo {title} {Coherence in spontaneous
  radiation processes},\ }\href
  {https://journals.aps.org/pr/abstract/10.1103/PhysRev.93.99} {\bibfield
  {journal} {\bibinfo  {journal} {Phys. Rev.}\ }\textbf {\bibinfo {volume}
  {93}},\ \bibinfo {pages} {99} (\bibinfo {year} {1954})}\BibitemShut {NoStop}%
\bibitem [{\citenamefont {Skribanowitz}\ \emph {et~al.}(1973)\citenamefont
  {Skribanowitz}, \citenamefont {Herman}, \citenamefont {MacGillivray},\ and\
  \citenamefont {Feld}}]{skribanowitz1973observation}%
  \BibitemOpen
  \bibfield  {author} {\bibinfo {author} {\bibfnamefont {N.}~\bibnamefont
  {Skribanowitz}}, \bibinfo {author} {\bibfnamefont {I.}~\bibnamefont
  {Herman}}, \bibinfo {author} {\bibfnamefont {J.}~\bibnamefont
  {MacGillivray}},\ and\ \bibinfo {author} {\bibfnamefont {M.}~\bibnamefont
  {Feld}},\ }\bibfield  {title} {\bibinfo {title} {Observation of {Dicke}
  superradiance in optically pumped {HF} gas},\ }\href
  {https://journals.aps.org/prl/abstract/10.1103/PhysRevLett.30.309} {\bibfield
   {journal} {\bibinfo  {journal} {Phys. Rev. Lett.}\ }\textbf {\bibinfo
  {volume} {30}},\ \bibinfo {pages} {309} (\bibinfo {year} {1973})}\BibitemShut
  {NoStop}%
\bibitem [{\citenamefont {Gross}\ \emph {et~al.}(1976)\citenamefont {Gross},
  \citenamefont {Fabre}, \citenamefont {Pillet},\ and\ \citenamefont
  {Haroche}}]{gross1976observation}%
  \BibitemOpen
  \bibfield  {author} {\bibinfo {author} {\bibfnamefont {M.}~\bibnamefont
  {Gross}}, \bibinfo {author} {\bibfnamefont {C.}~\bibnamefont {Fabre}},
  \bibinfo {author} {\bibfnamefont {P.}~\bibnamefont {Pillet}},\ and\ \bibinfo
  {author} {\bibfnamefont {S.}~\bibnamefont {Haroche}},\ }\bibfield  {title}
  {\bibinfo {title} {Observation of near-infrared {Dicke} superradiance on
  cascading transitions in atomic sodium},\ }\href
  {https://journals.aps.org/prl/abstract/10.1103/PhysRevLett.36.1035}
  {\bibfield  {journal} {\bibinfo  {journal} {Phys. Rev. Lett.}\ }\textbf
  {\bibinfo {volume} {36}},\ \bibinfo {pages} {1035} (\bibinfo {year}
  {1976})}\BibitemShut {NoStop}%
\bibitem [{\citenamefont {Gross}\ and\ \citenamefont
  {Haroche}(1982)}]{gross1982superradiance}%
  \BibitemOpen
  \bibfield  {author} {\bibinfo {author} {\bibfnamefont {M.}~\bibnamefont
  {Gross}}\ and\ \bibinfo {author} {\bibfnamefont {S.}~\bibnamefont
  {Haroche}},\ }\bibfield  {title} {\bibinfo {title} {Superradiance: An essay
  on the theory of collective spontaneous emission},\ }\href
  {https://www.sciencedirect.com/science/article/pii/0370157382901028}
  {\bibfield  {journal} {\bibinfo  {journal} {Phys. Rep.}\ }\textbf {\bibinfo
  {volume} {93}},\ \bibinfo {pages} {301} (\bibinfo {year} {1982})}\BibitemShut
  {NoStop}%
\bibitem [{\citenamefont {Cardenas-Lopez}\ \emph {et~al.}(2023)\citenamefont
  {Cardenas-Lopez}, \citenamefont {Masson}, \citenamefont {Zager},\ and\
  \citenamefont {Asenjo-Garcia}}]{cardenaslopez2023manybody}%
  \BibitemOpen
  \bibfield  {author} {\bibinfo {author} {\bibfnamefont {S.}~\bibnamefont
  {Cardenas-Lopez}}, \bibinfo {author} {\bibfnamefont {S.~J.}\ \bibnamefont
  {Masson}}, \bibinfo {author} {\bibfnamefont {Z.}~\bibnamefont {Zager}},\ and\
  \bibinfo {author} {\bibfnamefont {A.}~\bibnamefont {Asenjo-Garcia}},\
  }\bibfield  {title} {\bibinfo {title} {Many-body superradiance and dynamical
  mirror symmetry breaking in waveguide qed},\ }\href
  {https://doi.org/10.1103/PhysRevLett.131.033605} {\bibfield  {journal}
  {\bibinfo  {journal} {Phys. Rev. Lett.}\ }\textbf {\bibinfo {volume} {131}},\
  \bibinfo {pages} {033605} (\bibinfo {year} {2023})}\BibitemShut {NoStop}%
\bibitem [{\citenamefont {Liedl}\ \emph {et~al.}(2024)\citenamefont {Liedl},
  \citenamefont {Tebbenjohanns}, \citenamefont {Bach}, \citenamefont {Pucher},
  \citenamefont {Rauschenbeutel},\ and\ \citenamefont
  {Schneeweiss}}]{liedl2024observation}%
  \BibitemOpen
  \bibfield  {author} {\bibinfo {author} {\bibfnamefont {C.}~\bibnamefont
  {Liedl}}, \bibinfo {author} {\bibfnamefont {F.}~\bibnamefont
  {Tebbenjohanns}}, \bibinfo {author} {\bibfnamefont {C.}~\bibnamefont {Bach}},
  \bibinfo {author} {\bibfnamefont {S.}~\bibnamefont {Pucher}}, \bibinfo
  {author} {\bibfnamefont {A.}~\bibnamefont {Rauschenbeutel}},\ and\ \bibinfo
  {author} {\bibfnamefont {P.}~\bibnamefont {Schneeweiss}},\ }\bibfield
  {title} {\bibinfo {title} {Observation of superradiant bursts in a cascaded
  quantum system},\ }\href {https://doi.org/10.1103/PhysRevX.14.011020}
  {\bibfield  {journal} {\bibinfo  {journal} {Phys. Rev. X}\ }\textbf {\bibinfo
  {volume} {14}},\ \bibinfo {pages} {011020} (\bibinfo {year}
  {2024})}\BibitemShut {NoStop}%
\bibitem [{\citenamefont {Gardiner}(1993)}]{gardiner1993driving}%
  \BibitemOpen
  \bibfield  {author} {\bibinfo {author} {\bibfnamefont {C.~W.}\ \bibnamefont
  {Gardiner}},\ }\bibfield  {title} {\bibinfo {title} {Driving a quantum system
  with the output field from another driven quantum system},\ }\href
  {https://journals.aps.org/prl/abstract/10.1103/PhysRevLett.70.2269}
  {\bibfield  {journal} {\bibinfo  {journal} {Phys. Rev. Lett.}\ }\textbf
  {\bibinfo {volume} {70}},\ \bibinfo {pages} {2269} (\bibinfo {year}
  {1993})}\BibitemShut {NoStop}%
\bibitem [{\citenamefont {Carmichael}(1993)}]{carmichael1993quantum}%
  \BibitemOpen
  \bibfield  {author} {\bibinfo {author} {\bibfnamefont {H.~J.}\ \bibnamefont
  {Carmichael}},\ }\bibfield  {title} {\bibinfo {title} {Quantum trajectory
  theory for cascaded open systems},\ }\href
  {https://journals.aps.org/prl/abstract/10.1103/PhysRevLett.70.2273}
  {\bibfield  {journal} {\bibinfo  {journal} {Phys. Rev. Lett}\ }\textbf
  {\bibinfo {volume} {70}},\ \bibinfo {pages} {2273} (\bibinfo {year}
  {1993})}\BibitemShut {NoStop}%
\bibitem [{\citenamefont {Stiesdal}\ \emph {et~al.}(2021)\citenamefont
  {Stiesdal}, \citenamefont {Busche}, \citenamefont {Kleinbeck}, \citenamefont
  {Kumlin}, \citenamefont {Hansen}, \citenamefont {B{\"u}chler},\ and\
  \citenamefont {Hofferberth}}]{stiesdal2021controlled}%
  \BibitemOpen
  \bibfield  {author} {\bibinfo {author} {\bibfnamefont {N.}~\bibnamefont
  {Stiesdal}}, \bibinfo {author} {\bibfnamefont {H.}~\bibnamefont {Busche}},
  \bibinfo {author} {\bibfnamefont {K.}~\bibnamefont {Kleinbeck}}, \bibinfo
  {author} {\bibfnamefont {J.}~\bibnamefont {Kumlin}}, \bibinfo {author}
  {\bibfnamefont {M.~G.}\ \bibnamefont {Hansen}}, \bibinfo {author}
  {\bibfnamefont {H.~P.}\ \bibnamefont {B{\"u}chler}},\ and\ \bibinfo {author}
  {\bibfnamefont {S.}~\bibnamefont {Hofferberth}},\ }\bibfield  {title}
  {\bibinfo {title} {Controlled multi-photon subtraction with cascaded rydberg
  superatoms as single-photon absorbers},\ }\href
  {https://doi.org/https://doi.org/10.1038/s41467-021-24522-w} {\bibfield
  {journal} {\bibinfo  {journal} {Nat. Commun.}\ }\textbf {\bibinfo {volume}
  {12}},\ \bibinfo {pages} {4328} (\bibinfo {year} {2021})}\BibitemShut
  {NoStop}%
\bibitem [{\citenamefont {Calaj\'o}\ \emph {et~al.}(2019)\citenamefont
  {Calaj\'o}, \citenamefont {Schuetz}, \citenamefont {Pichler}, \citenamefont
  {Lukin}, \citenamefont {Schneeweiss}, \citenamefont {Volz},\ and\
  \citenamefont {Rabl}}]{calajo2019quantum}%
  \BibitemOpen
  \bibfield  {author} {\bibinfo {author} {\bibfnamefont {G.}~\bibnamefont
  {Calaj\'o}}, \bibinfo {author} {\bibfnamefont {M.~J.~A.}\ \bibnamefont
  {Schuetz}}, \bibinfo {author} {\bibfnamefont {H.}~\bibnamefont {Pichler}},
  \bibinfo {author} {\bibfnamefont {M.~D.}\ \bibnamefont {Lukin}}, \bibinfo
  {author} {\bibfnamefont {P.}~\bibnamefont {Schneeweiss}}, \bibinfo {author}
  {\bibfnamefont {J.}~\bibnamefont {Volz}},\ and\ \bibinfo {author}
  {\bibfnamefont {P.}~\bibnamefont {Rabl}},\ }\bibfield  {title} {\bibinfo
  {title} {Quantum acousto-optic control of light-matter interactions in
  nanophotonic networks},\ }\href {https://doi.org/10.1103/PhysRevA.99.053852}
  {\bibfield  {journal} {\bibinfo  {journal} {Phys. Rev. A}\ }\textbf {\bibinfo
  {volume} {99}},\ \bibinfo {pages} {053852} (\bibinfo {year}
  {2019})}\BibitemShut {NoStop}%
\bibitem [{\citenamefont {Ramos}\ \emph {et~al.}(2014)\citenamefont {Ramos},
  \citenamefont {Pichler}, \citenamefont {Daley},\ and\ \citenamefont
  {Zoller}}]{ramos2014quantum}%
  \BibitemOpen
  \bibfield  {author} {\bibinfo {author} {\bibfnamefont {T.}~\bibnamefont
  {Ramos}}, \bibinfo {author} {\bibfnamefont {H.}~\bibnamefont {Pichler}},
  \bibinfo {author} {\bibfnamefont {A.~J.}\ \bibnamefont {Daley}},\ and\
  \bibinfo {author} {\bibfnamefont {P.}~\bibnamefont {Zoller}},\ }\bibfield
  {title} {\bibinfo {title} {Quantum spin dimers from chiral dissipation in
  cold-atom chains},\ }\href {https://doi.org/10.1103/PhysRevLett.113.237203}
  {\bibfield  {journal} {\bibinfo  {journal} {Phys. Rev. Lett.}\ }\textbf
  {\bibinfo {volume} {113}},\ \bibinfo {pages} {237203} (\bibinfo {year}
  {2014})}\BibitemShut {NoStop}%
\bibitem [{\citenamefont {Lodahl}\ \emph {et~al.}(2017)\citenamefont {Lodahl},
  \citenamefont {Mahmoodian}, \citenamefont {Stobbe}, \citenamefont
  {Rauschenbeutel}, \citenamefont {Schneeweiss}, \citenamefont {Volz},
  \citenamefont {Pichler},\ and\ \citenamefont {Zoller}}]{lodahl2017chiral}%
  \BibitemOpen
  \bibfield  {author} {\bibinfo {author} {\bibfnamefont {P.}~\bibnamefont
  {Lodahl}}, \bibinfo {author} {\bibfnamefont {S.}~\bibnamefont {Mahmoodian}},
  \bibinfo {author} {\bibfnamefont {S.}~\bibnamefont {Stobbe}}, \bibinfo
  {author} {\bibfnamefont {A.}~\bibnamefont {Rauschenbeutel}}, \bibinfo
  {author} {\bibfnamefont {P.}~\bibnamefont {Schneeweiss}}, \bibinfo {author}
  {\bibfnamefont {J.}~\bibnamefont {Volz}}, \bibinfo {author} {\bibfnamefont
  {H.}~\bibnamefont {Pichler}},\ and\ \bibinfo {author} {\bibfnamefont
  {P.}~\bibnamefont {Zoller}},\ }\bibfield  {title} {\bibinfo {title} {Chiral
  quantum optics},\ }\href {https://doi.org/10.1038/nature21037} {\bibfield
  {journal} {\bibinfo  {journal} {Nature}\ }\textbf {\bibinfo {volume} {541}},\
  \bibinfo {pages} {473} (\bibinfo {year} {2017})}\BibitemShut {NoStop}%
\bibitem [{\citenamefont {Bohnet}\ \emph {et~al.}(2012)\citenamefont {Bohnet},
  \citenamefont {Chen}, \citenamefont {Weiner}, \citenamefont {Meiser},
  \citenamefont {Holland},\ and\ \citenamefont
  {Thompson}}]{bohnet2011steadystate}%
  \BibitemOpen
  \bibfield  {author} {\bibinfo {author} {\bibfnamefont {J.~G.}\ \bibnamefont
  {Bohnet}}, \bibinfo {author} {\bibfnamefont {Z.}~\bibnamefont {Chen}},
  \bibinfo {author} {\bibfnamefont {J.~M.}\ \bibnamefont {Weiner}}, \bibinfo
  {author} {\bibfnamefont {D.}~\bibnamefont {Meiser}}, \bibinfo {author}
  {\bibfnamefont {M.~J.}\ \bibnamefont {Holland}},\ and\ \bibinfo {author}
  {\bibfnamefont {J.~K.}\ \bibnamefont {Thompson}},\ }\bibfield  {title}
  {\bibinfo {title} {A steady-state superradiant laser with less than one
  intracavity photon},\ }\href {https://doi.org/10.1038/nature10920} {\bibfield
   {journal} {\bibinfo  {journal} {Nature}\ }\textbf {\bibinfo {volume}
  {484}},\ \bibinfo {pages} {78} (\bibinfo {year} {2012})}\BibitemShut
  {NoStop}%
\bibitem [{\citenamefont {Bienaimé}\ \emph {et~al.}(2013)\citenamefont
  {Bienaimé}, \citenamefont {Bachelard}, \citenamefont {Piovella},\ and\
  \citenamefont {Kaiser}}]{bienaime2013cooperativity}%
  \BibitemOpen
  \bibfield  {author} {\bibinfo {author} {\bibfnamefont {T.}~\bibnamefont
  {Bienaimé}}, \bibinfo {author} {\bibfnamefont {R.}~\bibnamefont
  {Bachelard}}, \bibinfo {author} {\bibfnamefont {N.}~\bibnamefont
  {Piovella}},\ and\ \bibinfo {author} {\bibfnamefont {R.}~\bibnamefont
  {Kaiser}},\ }\bibfield  {title} {\bibinfo {title} {Cooperativity in light
  scattering by cold atoms},\ }\href
  {https://doi.org/https://doi.org/10.1002/prop.201200089} {\bibfield
  {journal} {\bibinfo  {journal} {Fortschritte der Physik}\ }\textbf {\bibinfo
  {volume} {61}},\ \bibinfo {pages} {377} (\bibinfo {year} {2013})}\BibitemShut
  {NoStop}%
\bibitem [{\citenamefont {Cong}\ \emph {et~al.}(2016)\citenamefont {Cong},
  \citenamefont {Zhang}, \citenamefont {Wang}, \citenamefont {Noe},
  \citenamefont {Belyanin},\ and\ \citenamefont {Kono}}]{cong2016dicke}%
  \BibitemOpen
  \bibfield  {author} {\bibinfo {author} {\bibfnamefont {K.}~\bibnamefont
  {Cong}}, \bibinfo {author} {\bibfnamefont {Q.}~\bibnamefont {Zhang}},
  \bibinfo {author} {\bibfnamefont {Y.}~\bibnamefont {Wang}}, \bibinfo {author}
  {\bibfnamefont {G.~T.}\ \bibnamefont {Noe}}, \bibinfo {author} {\bibfnamefont
  {A.}~\bibnamefont {Belyanin}},\ and\ \bibinfo {author} {\bibfnamefont
  {J.}~\bibnamefont {Kono}},\ }\bibfield  {title} {\bibinfo {title} {Dicke
  superradiance in solids [invited]},\ }\href
  {https://doi.org/10.1364/JOSAB.33.000C80} {\bibfield  {journal} {\bibinfo
  {journal} {J. Opt. Soc. Am. B}\ }\textbf {\bibinfo {volume} {33}},\ \bibinfo
  {pages} {C80} (\bibinfo {year} {2016})}\BibitemShut {NoStop}%
\bibitem [{\citenamefont {Laske}\ \emph {et~al.}(2019)\citenamefont {Laske},
  \citenamefont {Winter},\ and\ \citenamefont {Hemmerich}}]{laske2019pulse}%
  \BibitemOpen
  \bibfield  {author} {\bibinfo {author} {\bibfnamefont {T.}~\bibnamefont
  {Laske}}, \bibinfo {author} {\bibfnamefont {H.}~\bibnamefont {Winter}},\ and\
  \bibinfo {author} {\bibfnamefont {A.}~\bibnamefont {Hemmerich}},\ }\bibfield
  {title} {\bibinfo {title} {Pulse delay time statistics in a superradiant
  laser with calcium atoms},\ }\href
  {https://doi.org/10.1103/PhysRevLett.123.103601} {\bibfield  {journal}
  {\bibinfo  {journal} {Phys. Rev. Lett.}\ }\textbf {\bibinfo {volume} {123}},\
  \bibinfo {pages} {103601} (\bibinfo {year} {2019})}\BibitemShut {NoStop}%
\bibitem [{\citenamefont {Ferioli}\ \emph {et~al.}(2021)\citenamefont
  {Ferioli}, \citenamefont {Glicenstein}, \citenamefont {Robicheaux},
  \citenamefont {Sutherland}, \citenamefont {Browaeys},\ and\ \citenamefont
  {Ferrier-Barbut}}]{ferioli2021laser}%
  \BibitemOpen
  \bibfield  {author} {\bibinfo {author} {\bibfnamefont {G.}~\bibnamefont
  {Ferioli}}, \bibinfo {author} {\bibfnamefont {A.}~\bibnamefont
  {Glicenstein}}, \bibinfo {author} {\bibfnamefont {F.}~\bibnamefont
  {Robicheaux}}, \bibinfo {author} {\bibfnamefont {R.~T.}\ \bibnamefont
  {Sutherland}}, \bibinfo {author} {\bibfnamefont {A.}~\bibnamefont
  {Browaeys}},\ and\ \bibinfo {author} {\bibfnamefont {I.}~\bibnamefont
  {Ferrier-Barbut}},\ }\bibfield  {title} {\bibinfo {title} {Laser-driven
  superradiant ensembles of two-level atoms near dicke regime},\ }\href
  {https://doi.org/10.1103/PhysRevLett.127.243602} {\bibfield  {journal}
  {\bibinfo  {journal} {Phys. Rev. Lett.}\ }\textbf {\bibinfo {volume} {127}},\
  \bibinfo {pages} {243602} (\bibinfo {year} {2021})}\BibitemShut {NoStop}%
\bibitem [{\citenamefont {Pennetta}\ \emph {et~al.}(2022)\citenamefont
  {Pennetta}, \citenamefont {Lechner}, \citenamefont {Blaha}, \citenamefont
  {Rauschenbeutel}, \citenamefont {Schneeweiss},\ and\ \citenamefont
  {Volz}}]{pennetta2022observation}%
  \BibitemOpen
  \bibfield  {author} {\bibinfo {author} {\bibfnamefont {R.}~\bibnamefont
  {Pennetta}}, \bibinfo {author} {\bibfnamefont {D.}~\bibnamefont {Lechner}},
  \bibinfo {author} {\bibfnamefont {M.}~\bibnamefont {Blaha}}, \bibinfo
  {author} {\bibfnamefont {A.}~\bibnamefont {Rauschenbeutel}}, \bibinfo
  {author} {\bibfnamefont {P.}~\bibnamefont {Schneeweiss}},\ and\ \bibinfo
  {author} {\bibfnamefont {J.}~\bibnamefont {Volz}},\ }\bibfield  {title}
  {\bibinfo {title} {Observation of coherent coupling between super- and
  subradiant states of an ensemble of cold atoms collectively coupled to a
  single propagating optical mode},\ }\href
  {https://doi.org/10.1103/PhysRevLett.128.203601} {\bibfield  {journal}
  {\bibinfo  {journal} {Phys. Rev. Lett.}\ }\textbf {\bibinfo {volume} {128}},\
  \bibinfo {pages} {203601} (\bibinfo {year} {2022})}\BibitemShut {NoStop}%
\bibitem [{\citenamefont {Reitz}\ \emph {et~al.}(2022)\citenamefont {Reitz},
  \citenamefont {Sommer},\ and\ \citenamefont {Genes}}]{reitz2022cooperative}%
  \BibitemOpen
  \bibfield  {author} {\bibinfo {author} {\bibfnamefont {M.}~\bibnamefont
  {Reitz}}, \bibinfo {author} {\bibfnamefont {C.}~\bibnamefont {Sommer}},\ and\
  \bibinfo {author} {\bibfnamefont {C.}~\bibnamefont {Genes}},\ }\bibfield
  {title} {\bibinfo {title} {Cooperative quantum phenomena in light-matter
  platforms},\ }\href {https://doi.org/10.1103/PRXQuantum.3.010201} {\bibfield
  {journal} {\bibinfo  {journal} {PRX Quantum}\ }\textbf {\bibinfo {volume}
  {3}},\ \bibinfo {pages} {010201} (\bibinfo {year} {2022})}\BibitemShut
  {NoStop}%
\bibitem [{\citenamefont {Santos}\ and\ \citenamefont
  {Bachelard}(2023)}]{santos2023generation}%
  \BibitemOpen
  \bibfield  {author} {\bibinfo {author} {\bibfnamefont {A.~C.}\ \bibnamefont
  {Santos}}\ and\ \bibinfo {author} {\bibfnamefont {R.}~\bibnamefont
  {Bachelard}},\ }\bibfield  {title} {\bibinfo {title} {Generation of maximally
  entangled long-lived states with giant atoms in a waveguide},\ }\href
  {https://doi.org/10.1103/PhysRevLett.130.053601} {\bibfield  {journal}
  {\bibinfo  {journal} {Phys. Rev. Lett.}\ }\textbf {\bibinfo {volume} {130}},\
  \bibinfo {pages} {053601} (\bibinfo {year} {2023})}\BibitemShut {NoStop}%
\bibitem [{\citenamefont {Ferioli}\ \emph {et~al.}(2024)\citenamefont
  {Ferioli}, \citenamefont {Pancaldi}, \citenamefont {Glicenstein},
  \citenamefont {Cl\'ement}, \citenamefont {Browaeys},\ and\ \citenamefont
  {Ferrier-Barbut}}]{ferioli2024non}%
  \BibitemOpen
  \bibfield  {author} {\bibinfo {author} {\bibfnamefont {G.}~\bibnamefont
  {Ferioli}}, \bibinfo {author} {\bibfnamefont {S.}~\bibnamefont {Pancaldi}},
  \bibinfo {author} {\bibfnamefont {A.}~\bibnamefont {Glicenstein}}, \bibinfo
  {author} {\bibfnamefont {D.}~\bibnamefont {Cl\'ement}}, \bibinfo {author}
  {\bibfnamefont {A.}~\bibnamefont {Browaeys}},\ and\ \bibinfo {author}
  {\bibfnamefont {I.}~\bibnamefont {Ferrier-Barbut}},\ }\bibfield  {title}
  {\bibinfo {title} {Non-gaussian correlations in the steady state of
  driven-dissipative clouds of two-level atoms},\ }\href
  {https://doi.org/10.1103/PhysRevLett.132.133601} {\bibfield  {journal}
  {\bibinfo  {journal} {Phys. Rev. Lett.}\ }\textbf {\bibinfo {volume} {132}},\
  \bibinfo {pages} {133601} (\bibinfo {year} {2024})}\BibitemShut {NoStop}%
\bibitem [{\citenamefont {Ostermann}\ \emph {et~al.}(2024)\citenamefont
  {Ostermann}, \citenamefont {Rubies-Bigorda}, \citenamefont {Zhang},\ and\
  \citenamefont {Yelin}}]{ostermann2024breakdown}%
  \BibitemOpen
  \bibfield  {author} {\bibinfo {author} {\bibfnamefont {S.}~\bibnamefont
  {Ostermann}}, \bibinfo {author} {\bibfnamefont {O.}~\bibnamefont
  {Rubies-Bigorda}}, \bibinfo {author} {\bibfnamefont {V.}~\bibnamefont
  {Zhang}},\ and\ \bibinfo {author} {\bibfnamefont {S.~F.}\ \bibnamefont
  {Yelin}},\ }\bibfield  {title} {\bibinfo {title} {Breakdown of steady-state
  superradiance in extended driven atomic arrays},\ }\href
  {https://doi.org/10.1103/PhysRevResearch.6.023206} {\bibfield  {journal}
  {\bibinfo  {journal} {Phys. Rev. Res.}\ }\textbf {\bibinfo {volume} {6}},\
  \bibinfo {pages} {023206} (\bibinfo {year} {2024})}\BibitemShut {NoStop}%
\bibitem [{\citenamefont {Haake}\ and\ \citenamefont
  {Glauber}(1972)}]{haake1972quantum}%
  \BibitemOpen
  \bibfield  {author} {\bibinfo {author} {\bibfnamefont {F.}~\bibnamefont
  {Haake}}\ and\ \bibinfo {author} {\bibfnamefont {R.~J.}\ \bibnamefont
  {Glauber}},\ }\bibfield  {title} {\bibinfo {title} {Quantum statistics of
  superradiant pulses},\ }\href {https://doi.org/10.1103/PhysRevA.5.1457}
  {\bibfield  {journal} {\bibinfo  {journal} {Phys. Rev. A}\ }\textbf {\bibinfo
  {volume} {5}},\ \bibinfo {pages} {1457} (\bibinfo {year} {1972})}\BibitemShut
  {NoStop}%
\bibitem [{\citenamefont {Lopes}\ \emph {et~al.}(2014)\citenamefont {Lopes},
  \citenamefont {Imanaliev}, \citenamefont {Bonneau}, \citenamefont {Ruaudel},
  \citenamefont {Cheneau}, \citenamefont {Boiron},\ and\ \citenamefont
  {Westbrook}}]{lopes2014secondorder}%
  \BibitemOpen
  \bibfield  {author} {\bibinfo {author} {\bibfnamefont {R.}~\bibnamefont
  {Lopes}}, \bibinfo {author} {\bibfnamefont {A.}~\bibnamefont {Imanaliev}},
  \bibinfo {author} {\bibfnamefont {M.}~\bibnamefont {Bonneau}}, \bibinfo
  {author} {\bibfnamefont {J.}~\bibnamefont {Ruaudel}}, \bibinfo {author}
  {\bibfnamefont {M.}~\bibnamefont {Cheneau}}, \bibinfo {author} {\bibfnamefont
  {D.}~\bibnamefont {Boiron}},\ and\ \bibinfo {author} {\bibfnamefont {C.~I.}\
  \bibnamefont {Westbrook}},\ }\bibfield  {title} {\bibinfo {title}
  {Second-order coherence of superradiance from a bose-einstein condensate},\
  }\href {https://doi.org/10.1103/PhysRevA.90.013615} {\bibfield  {journal}
  {\bibinfo  {journal} {Phys. Rev. A}\ }\textbf {\bibinfo {volume} {90}},\
  \bibinfo {pages} {013615} (\bibinfo {year} {2014})}\BibitemShut {NoStop}%
\bibitem [{\citenamefont {Jahnke}\ \emph {et~al.}(2016)\citenamefont {Jahnke},
  \citenamefont {Gies}, \citenamefont {A{\ss}mann}, \citenamefont {Bayer},
  \citenamefont {Leymann}, \citenamefont {Foerster}, \citenamefont {Wiersig},
  \citenamefont {Schneider}, \citenamefont {Kamp},\ and\ \citenamefont
  {H{\"o}fling}}]{jahnke2016giant}%
  \BibitemOpen
  \bibfield  {author} {\bibinfo {author} {\bibfnamefont {F.}~\bibnamefont
  {Jahnke}}, \bibinfo {author} {\bibfnamefont {C.}~\bibnamefont {Gies}},
  \bibinfo {author} {\bibfnamefont {M.}~\bibnamefont {A{\ss}mann}}, \bibinfo
  {author} {\bibfnamefont {M.}~\bibnamefont {Bayer}}, \bibinfo {author}
  {\bibfnamefont {H.}~\bibnamefont {Leymann}}, \bibinfo {author} {\bibfnamefont
  {A.}~\bibnamefont {Foerster}}, \bibinfo {author} {\bibfnamefont
  {J.}~\bibnamefont {Wiersig}}, \bibinfo {author} {\bibfnamefont
  {C.}~\bibnamefont {Schneider}}, \bibinfo {author} {\bibfnamefont
  {M.}~\bibnamefont {Kamp}},\ and\ \bibinfo {author} {\bibfnamefont
  {S.}~\bibnamefont {H{\"o}fling}},\ }\bibfield  {title} {\bibinfo {title}
  {Giant photon bunching, superradiant pulse emission and excitation trapping
  in quantum-dot nanolasers},\ }\href
  {https://www.nature.com/articles/ncomms11540} {\bibfield  {journal} {\bibinfo
   {journal} {Nat. Commun.}\ }\textbf {\bibinfo {volume} {7}},\ \bibinfo
  {pages} {11540} (\bibinfo {year} {2016})}\BibitemShut {NoStop}%
\bibitem [{\citenamefont {Stryzhenko}\ \emph {et~al.}(2024)\citenamefont
  {Stryzhenko}, \citenamefont {Bruns},\ and\ \citenamefont
  {Peters}}]{stryzhenko2024Nscaling}%
  \BibitemOpen
  \bibfield  {author} {\bibinfo {author} {\bibfnamefont {S.}~\bibnamefont
  {Stryzhenko}}, \bibinfo {author} {\bibfnamefont {A.}~\bibnamefont {Bruns}},\
  and\ \bibinfo {author} {\bibfnamefont {T.}~\bibnamefont {Peters}},\
  }\bibfield  {title} {\bibinfo {title} {$n$ scaling of large-sample collective
  decay in inhomogeneous ensembles},\ }\href
  {https://doi.org/10.1103/PhysRevResearch.6.013091} {\bibfield  {journal}
  {\bibinfo  {journal} {Phys. Rev. Res.}\ }\textbf {\bibinfo {volume} {6}},\
  \bibinfo {pages} {013091} (\bibinfo {year} {2024})}\BibitemShut {NoStop}%
\bibitem [{\citenamefont {Glauber}(1963)}]{glauber1963quantum}%
  \BibitemOpen
  \bibfield  {author} {\bibinfo {author} {\bibfnamefont {R.~J.}\ \bibnamefont
  {Glauber}},\ }\bibfield  {title} {\bibinfo {title} {The quantum theory of
  optical coherence},\ }\href {https://doi.org/10.1103/PhysRev.130.2529}
  {\bibfield  {journal} {\bibinfo  {journal} {Phys. Rev.}\ }\textbf {\bibinfo
  {volume} {130}},\ \bibinfo {pages} {2529} (\bibinfo {year}
  {1963})}\BibitemShut {NoStop}%
\bibitem [{\citenamefont {Vetsch}\ \emph {et~al.}(2010)\citenamefont {Vetsch},
  \citenamefont {Reitz}, \citenamefont {Sagu{\'e}}, \citenamefont {Schmidt},
  \citenamefont {Dawkins},\ and\ \citenamefont
  {Rauschenbeutel}}]{vetsch2010optical}%
  \BibitemOpen
  \bibfield  {author} {\bibinfo {author} {\bibfnamefont {E.}~\bibnamefont
  {Vetsch}}, \bibinfo {author} {\bibfnamefont {D.}~\bibnamefont {Reitz}},
  \bibinfo {author} {\bibfnamefont {G.}~\bibnamefont {Sagu{\'e}}}, \bibinfo
  {author} {\bibfnamefont {R.}~\bibnamefont {Schmidt}}, \bibinfo {author}
  {\bibfnamefont {S.}~\bibnamefont {Dawkins}},\ and\ \bibinfo {author}
  {\bibfnamefont {A.}~\bibnamefont {Rauschenbeutel}},\ }\bibfield  {title}
  {\bibinfo {title} {Optical interface created by laser-cooled atoms trapped in
  the evanescent field surrounding an optical nanofiber},\ }\href
  {https://doi.org/10.1103/PhysRevLett.104.203603} {\bibfield  {journal}
  {\bibinfo  {journal} {Phys. Rev. Lett.}\ }\textbf {\bibinfo {volume} {104}},\
  \bibinfo {pages} {203603} (\bibinfo {year} {2010})}\BibitemShut {NoStop}%
\bibitem [{\citenamefont {Schlosser}\ \emph {et~al.}(2002)\citenamefont
  {Schlosser}, \citenamefont {Reymond},\ and\ \citenamefont
  {Grangier}}]{Schlosser2002}%
  \BibitemOpen
  \bibfield  {author} {\bibinfo {author} {\bibfnamefont {N.}~\bibnamefont
  {Schlosser}}, \bibinfo {author} {\bibfnamefont {G.}~\bibnamefont {Reymond}},\
  and\ \bibinfo {author} {\bibfnamefont {P.}~\bibnamefont {Grangier}},\
  }\bibfield  {title} {\bibinfo {title} {Collisional blockade in microscopic
  optical dipole traps},\ }\href
  {https://doi.org/10.1103/PhysRevLett.89.023005} {\bibfield  {journal}
  {\bibinfo  {journal} {Phys. Rev. Lett.}\ }\textbf {\bibinfo {volume} {89}},\
  \bibinfo {pages} {023005} (\bibinfo {year} {2002})}\BibitemShut {NoStop}%
\bibitem [{\citenamefont {Meng}\ \emph {et~al.}(2018)\citenamefont {Meng},
  \citenamefont {Dareau}, \citenamefont {Schneeweiss},\ and\ \citenamefont
  {Rauschenbeutel}}]{Meng2018}%
  \BibitemOpen
  \bibfield  {author} {\bibinfo {author} {\bibfnamefont {Y.}~\bibnamefont
  {Meng}}, \bibinfo {author} {\bibfnamefont {A.}~\bibnamefont {Dareau}},
  \bibinfo {author} {\bibfnamefont {P.}~\bibnamefont {Schneeweiss}},\ and\
  \bibinfo {author} {\bibfnamefont {A.}~\bibnamefont {Rauschenbeutel}},\
  }\bibfield  {title} {\bibinfo {title} {Near-ground-state cooling of atoms
  optically trapped 300 nm away from a hot surface},\ }\href
  {https://doi.org/10.1103/PhysRevX.8.031054} {\bibfield  {journal} {\bibinfo
  {journal} {Phys. Rev. X}\ }\textbf {\bibinfo {volume} {8}},\ \bibinfo {pages}
  {031054} (\bibinfo {year} {2018})}\BibitemShut {NoStop}%
\bibitem [{\citenamefont {Mitsch}\ \emph {et~al.}(2014)\citenamefont {Mitsch},
  \citenamefont {Sayrin}, \citenamefont {Albrecht}, \citenamefont
  {Schneeweiss},\ and\ \citenamefont {Rauschenbeutel}}]{mitsch2014quantum}%
  \BibitemOpen
  \bibfield  {author} {\bibinfo {author} {\bibfnamefont {R.}~\bibnamefont
  {Mitsch}}, \bibinfo {author} {\bibfnamefont {C.}~\bibnamefont {Sayrin}},
  \bibinfo {author} {\bibfnamefont {B.}~\bibnamefont {Albrecht}}, \bibinfo
  {author} {\bibfnamefont {P.}~\bibnamefont {Schneeweiss}},\ and\ \bibinfo
  {author} {\bibfnamefont {A.}~\bibnamefont {Rauschenbeutel}},\ }\bibfield
  {title} {\bibinfo {title} {Quantum state-controlled directional spontaneous
  emission of photons into a nanophotonic waveguide},\ }\href
  {https://doi.org/10.1038/ncomms6713} {\bibfield  {journal} {\bibinfo
  {journal} {Nat. Commun.}\ }\textbf {\bibinfo {volume} {5}},\ \bibinfo {pages}
  {5713} (\bibinfo {year} {2014})}\BibitemShut {NoStop}%
\bibitem [{\citenamefont {Liedl}\ \emph {et~al.}(2023)\citenamefont {Liedl},
  \citenamefont {Pucher}, \citenamefont {Tebbenjohanns}, \citenamefont
  {Schneeweiss},\ and\ \citenamefont {Rauschenbeutel}}]{liedl2023collective}%
  \BibitemOpen
  \bibfield  {author} {\bibinfo {author} {\bibfnamefont {C.}~\bibnamefont
  {Liedl}}, \bibinfo {author} {\bibfnamefont {S.}~\bibnamefont {Pucher}},
  \bibinfo {author} {\bibfnamefont {F.}~\bibnamefont {Tebbenjohanns}}, \bibinfo
  {author} {\bibfnamefont {P.}~\bibnamefont {Schneeweiss}},\ and\ \bibinfo
  {author} {\bibfnamefont {A.}~\bibnamefont {Rauschenbeutel}},\ }\bibfield
  {title} {\bibinfo {title} {Collective radiation of a cascaded quantum system:
  From timed {Dicke} states to inverted ensembles},\ }\href
  {https://doi.org/10.1103/PhysRevLett.130.163602} {\bibfield  {journal}
  {\bibinfo  {journal} {Phys. Rev. Lett.}\ }\textbf {\bibinfo {volume} {130}},\
  \bibinfo {pages} {163602} (\bibinfo {year} {2023})}\BibitemShut {NoStop}%
\bibitem [{\citenamefont {Mink}\ and\ \citenamefont
  {Fleischhauer}(2023)}]{mink2023collective}%
  \BibitemOpen
  \bibfield  {author} {\bibinfo {author} {\bibfnamefont {C.~D.}\ \bibnamefont
  {Mink}}\ and\ \bibinfo {author} {\bibfnamefont {M.}~\bibnamefont
  {Fleischhauer}},\ }\bibfield  {title} {\bibinfo {title} {{Collective
  radiative interactions in the discrete truncated Wigner approximation}},\
  }\href {https://doi.org/10.21468/SciPostPhys.15.6.233} {\bibfield  {journal}
  {\bibinfo  {journal} {SciPost Phys.}\ }\textbf {\bibinfo {volume} {15}},\
  \bibinfo {pages} {233} (\bibinfo {year} {2023})}\BibitemShut {NoStop}%
\bibitem [{\citenamefont {Tebbenjohanns}\ \emph {et~al.}(2024)\citenamefont
  {Tebbenjohanns}, \citenamefont {Mink}, \citenamefont {Bach}, \citenamefont
  {Rauschenbeutel},\ and\ \citenamefont
  {Fleischhauer}}]{tebbenjohanns2024predicting}%
  \BibitemOpen
  \bibfield  {author} {\bibinfo {author} {\bibfnamefont {F.}~\bibnamefont
  {Tebbenjohanns}}, \bibinfo {author} {\bibfnamefont {C.~D.}\ \bibnamefont
  {Mink}}, \bibinfo {author} {\bibfnamefont {C.}~\bibnamefont {Bach}}, \bibinfo
  {author} {\bibfnamefont {A.}~\bibnamefont {Rauschenbeutel}},\ and\ \bibinfo
  {author} {\bibfnamefont {M.}~\bibnamefont {Fleischhauer}},\ }\href
  {https://arxiv.org/abs/2407.02154} {\bibinfo {title} {Predicting correlations
  in superradiant emission from a cascaded quantum system}} (\bibinfo {year}
  {2024}),\ \Eprint {https://arxiv.org/abs/2407.02154} {arXiv:2407.02154}
  \BibitemShut {NoStop}%
\bibitem [{\citenamefont {Ostermann}\ \emph {et~al.}(2012)\citenamefont
  {Ostermann}, \citenamefont {Zoubi},\ and\ \citenamefont
  {Ritsch}}]{ostermann2012collective}%
  \BibitemOpen
  \bibfield  {author} {\bibinfo {author} {\bibfnamefont {L.}~\bibnamefont
  {Ostermann}}, \bibinfo {author} {\bibfnamefont {H.}~\bibnamefont {Zoubi}},\
  and\ \bibinfo {author} {\bibfnamefont {H.}~\bibnamefont {Ritsch}},\
  }\bibfield  {title} {\bibinfo {title} {Cascaded collective decay in regular
  arrays of cold trapped atoms},\ }\href {https://doi.org/10.1364/OE.20.029634}
  {\bibfield  {journal} {\bibinfo  {journal} {Opt. Express}\ }\textbf {\bibinfo
  {volume} {20}},\ \bibinfo {pages} {29634} (\bibinfo {year}
  {2012})}\BibitemShut {NoStop}%
\bibitem [{\citenamefont {Caneva}\ \emph {et~al.}(2015)\citenamefont {Caneva},
  \citenamefont {Manzoni}, \citenamefont {Shi}, \citenamefont {Douglas},
  \citenamefont {Cirac},\ and\ \citenamefont
  {Chang}}]{caneva2015quantumdynamics}%
  \BibitemOpen
  \bibfield  {author} {\bibinfo {author} {\bibfnamefont {T.}~\bibnamefont
  {Caneva}}, \bibinfo {author} {\bibfnamefont {M.~T.}\ \bibnamefont {Manzoni}},
  \bibinfo {author} {\bibfnamefont {T.}~\bibnamefont {Shi}}, \bibinfo {author}
  {\bibfnamefont {J.~S.}\ \bibnamefont {Douglas}}, \bibinfo {author}
  {\bibfnamefont {J.~I.}\ \bibnamefont {Cirac}},\ and\ \bibinfo {author}
  {\bibfnamefont {D.~E.}\ \bibnamefont {Chang}},\ }\bibfield  {title} {\bibinfo
  {title} {Quantum dynamics of propagating photons with strong interactions: a
  generalized input–output formalism},\ }\href
  {https://doi.org/10.1088/1367-2630/17/11/113001} {\bibfield  {journal}
  {\bibinfo  {journal} {New Journal of Physics}\ }\textbf {\bibinfo {volume}
  {17}},\ \bibinfo {pages} {113001} (\bibinfo {year} {2015})}\BibitemShut
  {NoStop}%
\bibitem [{\citenamefont {Mahmoodian}\ \emph {et~al.}(2020)\citenamefont
  {Mahmoodian}, \citenamefont {Calaj{\'o}}, \citenamefont {Chang},
  \citenamefont {Hammerer},\ and\ \citenamefont
  {S{\o}rensen}}]{mahmoodian2020dynamics}%
  \BibitemOpen
  \bibfield  {author} {\bibinfo {author} {\bibfnamefont {S.}~\bibnamefont
  {Mahmoodian}}, \bibinfo {author} {\bibfnamefont {G.}~\bibnamefont
  {Calaj{\'o}}}, \bibinfo {author} {\bibfnamefont {D.~E.}\ \bibnamefont
  {Chang}}, \bibinfo {author} {\bibfnamefont {K.}~\bibnamefont {Hammerer}},\
  and\ \bibinfo {author} {\bibfnamefont {A.~S.}\ \bibnamefont {S{\o}rensen}},\
  }\bibfield  {title} {\bibinfo {title} {Dynamics of many-body photon bound
  states in chiral waveguide {QED}},\ }\href
  {https://journals.aps.org/prx/abstract/10.1103/PhysRevX.10.031011} {\bibfield
   {journal} {\bibinfo  {journal} {Phys. Rev. X}\ }\textbf {\bibinfo {volume}
  {10}},\ \bibinfo {pages} {031011} (\bibinfo {year} {2020})}\BibitemShut
  {NoStop}%
\bibitem [{\citenamefont {Robicheaux}\ and\ \citenamefont
  {Suresh}(2021)}]{robicheaux2021beyond}%
  \BibitemOpen
  \bibfield  {author} {\bibinfo {author} {\bibfnamefont {F.}~\bibnamefont
  {Robicheaux}}\ and\ \bibinfo {author} {\bibfnamefont {D.~A.}\ \bibnamefont
  {Suresh}},\ }\bibfield  {title} {\bibinfo {title} {Beyond lowest order
  mean-field theory for light interacting with atom arrays},\ }\href
  {https://doi.org/10.1103/PhysRevA.104.023702} {\bibfield  {journal} {\bibinfo
   {journal} {Phys. Rev. A}\ }\textbf {\bibinfo {volume} {104}},\ \bibinfo
  {pages} {023702} (\bibinfo {year} {2021})}\BibitemShut {NoStop}%
\bibitem [{\citenamefont {Arranz~Regidor}\ \emph {et~al.}(2021)\citenamefont
  {Arranz~Regidor}, \citenamefont {Crowder}, \citenamefont {Carmichael},\ and\
  \citenamefont {Hughes}}]{arranzregidor2021modeling}%
  \BibitemOpen
  \bibfield  {author} {\bibinfo {author} {\bibfnamefont {S.}~\bibnamefont
  {Arranz~Regidor}}, \bibinfo {author} {\bibfnamefont {G.}~\bibnamefont
  {Crowder}}, \bibinfo {author} {\bibfnamefont {H.}~\bibnamefont
  {Carmichael}},\ and\ \bibinfo {author} {\bibfnamefont {S.}~\bibnamefont
  {Hughes}},\ }\bibfield  {title} {\bibinfo {title} {Modeling quantum
  light-matter interactions in waveguide qed with retardation, nonlinear
  interactions, and a time-delayed feedback: Matrix product states versus a
  space-discretized waveguide model},\ }\href
  {https://doi.org/10.1103/PhysRevResearch.3.023030} {\bibfield  {journal}
  {\bibinfo  {journal} {Phys. Rev. Res.}\ }\textbf {\bibinfo {volume} {3}},\
  \bibinfo {pages} {023030} (\bibinfo {year} {2021})}\BibitemShut {NoStop}%
\bibitem [{\citenamefont {Kusmierek}\ \emph {et~al.}(2023)\citenamefont
  {Kusmierek}, \citenamefont {Mahmoodian}, \citenamefont {Cordier},
  \citenamefont {Hinney}, \citenamefont {Rauschenbeutel}, \citenamefont
  {Schemmer}, \citenamefont {Schneeweiss}, \citenamefont {Volz},\ and\
  \citenamefont {Hammerer}}]{kusmierek2023higherorder}%
  \BibitemOpen
  \bibfield  {author} {\bibinfo {author} {\bibfnamefont {K.~J.}\ \bibnamefont
  {Kusmierek}}, \bibinfo {author} {\bibfnamefont {S.}~\bibnamefont
  {Mahmoodian}}, \bibinfo {author} {\bibfnamefont {M.}~\bibnamefont {Cordier}},
  \bibinfo {author} {\bibfnamefont {J.}~\bibnamefont {Hinney}}, \bibinfo
  {author} {\bibfnamefont {A.}~\bibnamefont {Rauschenbeutel}}, \bibinfo
  {author} {\bibfnamefont {M.}~\bibnamefont {Schemmer}}, \bibinfo {author}
  {\bibfnamefont {P.}~\bibnamefont {Schneeweiss}}, \bibinfo {author}
  {\bibfnamefont {J.}~\bibnamefont {Volz}},\ and\ \bibinfo {author}
  {\bibfnamefont {K.}~\bibnamefont {Hammerer}},\ }\bibfield  {title} {\bibinfo
  {title} {{Higher-order mean-field theory of chiral waveguide {QED}}},\ }\href
  {https://doi.org/10.21468/SciPostPhysCore.6.2.041} {\bibfield  {journal}
  {\bibinfo  {journal} {SciPost Phys. Core}\ }\textbf {\bibinfo {volume} {6}},\
  \bibinfo {pages} {041} (\bibinfo {year} {2023})}\BibitemShut {NoStop}%
\bibitem [{\citenamefont {Rubies-Bigorda}\ \emph {et~al.}(2023)\citenamefont
  {Rubies-Bigorda}, \citenamefont {Ostermann},\ and\ \citenamefont
  {Yelin}}]{rubiesbigorda2023dynamic}%
  \BibitemOpen
  \bibfield  {author} {\bibinfo {author} {\bibfnamefont {O.}~\bibnamefont
  {Rubies-Bigorda}}, \bibinfo {author} {\bibfnamefont {S.}~\bibnamefont
  {Ostermann}},\ and\ \bibinfo {author} {\bibfnamefont {S.~F.}\ \bibnamefont
  {Yelin}},\ }\bibfield  {title} {\bibinfo {title} {Dynamic population of
  multiexcitation subradiant states in incoherently excited atomic arrays},\
  }\href {https://doi.org/10.1103/PhysRevA.107.L051701} {\bibfield  {journal}
  {\bibinfo  {journal} {Phys. Rev. A}\ }\textbf {\bibinfo {volume} {107}},\
  \bibinfo {pages} {L051701} (\bibinfo {year} {2023})}\BibitemShut {NoStop}%
\bibitem [{\citenamefont {Kleinbeck}\ \emph {et~al.}(2023)\citenamefont
  {Kleinbeck}, \citenamefont {Busche}, \citenamefont {Stiesdal}, \citenamefont
  {Hofferberth}, \citenamefont {M\o{}lmer},\ and\ \citenamefont
  {B\"uchler}}]{kleinbeck2023creation}%
  \BibitemOpen
  \bibfield  {author} {\bibinfo {author} {\bibfnamefont {K.}~\bibnamefont
  {Kleinbeck}}, \bibinfo {author} {\bibfnamefont {H.}~\bibnamefont {Busche}},
  \bibinfo {author} {\bibfnamefont {N.}~\bibnamefont {Stiesdal}}, \bibinfo
  {author} {\bibfnamefont {S.}~\bibnamefont {Hofferberth}}, \bibinfo {author}
  {\bibfnamefont {K.}~\bibnamefont {M\o{}lmer}},\ and\ \bibinfo {author}
  {\bibfnamefont {H.~P.}\ \bibnamefont {B\"uchler}},\ }\bibfield  {title}
  {\bibinfo {title} {Creation of nonclassical states of light in a chiral
  waveguide},\ }\href {https://doi.org/10.1103/PhysRevA.107.013717} {\bibfield
  {journal} {\bibinfo  {journal} {Phys. Rev. A}\ }\textbf {\bibinfo {volume}
  {107}},\ \bibinfo {pages} {013717} (\bibinfo {year} {2023})}\BibitemShut
  {NoStop}%
\bibitem [{\citenamefont {Vlasiuk}\ \emph {et~al.}(2023)\citenamefont
  {Vlasiuk}, \citenamefont {Poshakinskiy},\ and\ \citenamefont
  {Poddubny}}]{vlasiuk2023two}%
  \BibitemOpen
  \bibfield  {author} {\bibinfo {author} {\bibfnamefont {E.}~\bibnamefont
  {Vlasiuk}}, \bibinfo {author} {\bibfnamefont {A.~V.}\ \bibnamefont
  {Poshakinskiy}},\ and\ \bibinfo {author} {\bibfnamefont {A.~N.}\ \bibnamefont
  {Poddubny}},\ }\bibfield  {title} {\bibinfo {title} {Two-photon
  pulse-scattering spectroscopy for arrays of two-level atoms coupled to a
  waveguide},\ }\href {https://doi.org/10.1103/PhysRevA.108.033705} {\bibfield
  {journal} {\bibinfo  {journal} {Phys. Rev. A}\ }\textbf {\bibinfo {volume}
  {108}},\ \bibinfo {pages} {033705} (\bibinfo {year} {2023})}\BibitemShut
  {NoStop}%
\bibitem [{\citenamefont {Schachenmayer}\ \emph {et~al.}(2015)\citenamefont
  {Schachenmayer}, \citenamefont {Pikovski},\ and\ \citenamefont
  {Rey}}]{schachenmayer2015manybody}%
  \BibitemOpen
  \bibfield  {author} {\bibinfo {author} {\bibfnamefont {J.}~\bibnamefont
  {Schachenmayer}}, \bibinfo {author} {\bibfnamefont {A.}~\bibnamefont
  {Pikovski}},\ and\ \bibinfo {author} {\bibfnamefont {A.~M.}\ \bibnamefont
  {Rey}},\ }\bibfield  {title} {\bibinfo {title} {Many-body quantum spin
  dynamics with monte carlo trajectories on a discrete phase space},\ }\href
  {https://doi.org/10.1103/PhysRevX.5.011022} {\bibfield  {journal} {\bibinfo
  {journal} {Phys. Rev. X}\ }\textbf {\bibinfo {volume} {5}},\ \bibinfo {pages}
  {011022} (\bibinfo {year} {2015})}\BibitemShut {NoStop}%
\bibitem [{\citenamefont {Vrehen}\ \emph {et~al.}(1980)\citenamefont {Vrehen},
  \citenamefont {Schuurmans},\ and\ \citenamefont
  {Polder}}]{vrehen1980superfluorescence}%
  \BibitemOpen
  \bibfield  {author} {\bibinfo {author} {\bibfnamefont {Q.}~\bibnamefont
  {Vrehen}}, \bibinfo {author} {\bibfnamefont {M.}~\bibnamefont {Schuurmans}},\
  and\ \bibinfo {author} {\bibfnamefont {D.}~\bibnamefont {Polder}},\
  }\bibfield  {title} {\bibinfo {title} {Superfluorescence: macroscopic quantum
  fluctuations in the time domain},\ }\href {https://doi.org/10.1038/285070a0}
  {\bibfield  {journal} {\bibinfo  {journal} {Nature}\ }\textbf {\bibinfo
  {volume} {285}},\ \bibinfo {pages} {70} (\bibinfo {year} {1980})}\BibitemShut
  {NoStop}%
\bibitem [{\citenamefont {Bonifacio}\ and\ \citenamefont
  {Lugiato}(1975)}]{bonifacio1975cooperative}%
  \BibitemOpen
  \bibfield  {author} {\bibinfo {author} {\bibfnamefont {R.}~\bibnamefont
  {Bonifacio}}\ and\ \bibinfo {author} {\bibfnamefont {L.~A.}\ \bibnamefont
  {Lugiato}},\ }\bibfield  {title} {\bibinfo {title} {Cooperative radiation
  processes in two-level systems: Superfluorescence},\ }\href
  {https://doi.org/10.1103/PhysRevA.11.1507} {\bibfield  {journal} {\bibinfo
  {journal} {Phys. Rev. A}\ }\textbf {\bibinfo {volume} {11}},\ \bibinfo
  {pages} {1507} (\bibinfo {year} {1975})}\BibitemShut {NoStop}%
\bibitem [{\citenamefont {Ferreira}\ \emph {et~al.}(2020)\citenamefont
  {Ferreira}, \citenamefont {Bachelard}, \citenamefont {Guerin}, \citenamefont
  {Kaiser},\ and\ \citenamefont {Fouché}}]{ferreira2020connecting}%
  \BibitemOpen
  \bibfield  {author} {\bibinfo {author} {\bibfnamefont {D.}~\bibnamefont
  {Ferreira}}, \bibinfo {author} {\bibfnamefont {R.}~\bibnamefont {Bachelard}},
  \bibinfo {author} {\bibfnamefont {W.}~\bibnamefont {Guerin}}, \bibinfo
  {author} {\bibfnamefont {R.}~\bibnamefont {Kaiser}},\ and\ \bibinfo {author}
  {\bibfnamefont {M.}~\bibnamefont {Fouché}},\ }\bibfield  {title} {\bibinfo
  {title} {{Connecting field and intensity correlations: The Siegert relation
  and how to test it}},\ }\href {https://doi.org/10.1119/10.0001630} {\bibfield
   {journal} {\bibinfo  {journal} {American Journal of Physics}\ }\textbf
  {\bibinfo {volume} {88}},\ \bibinfo {pages} {831} (\bibinfo {year}
  {2020})}\BibitemShut {NoStop}%
\bibitem [{Note1()}]{Note1}%
  \BibitemOpen
  \bibinfo {note} {The normalized, equal-time first-order coherence function is
  by definition equal to unity, $g^{(1)}(t,t)=1$~\cite {glauber1963quantum,
  loudon2000quantum}}\BibitemShut {NoStop}%
\bibitem [{\citenamefont {Florian}\ \emph {et~al.}(1984)\citenamefont
  {Florian}, \citenamefont {Schwan},\ and\ \citenamefont
  {Schmid}}]{florian1984time}%
  \BibitemOpen
  \bibfield  {author} {\bibinfo {author} {\bibfnamefont {R.}~\bibnamefont
  {Florian}}, \bibinfo {author} {\bibfnamefont {L.~O.}\ \bibnamefont
  {Schwan}},\ and\ \bibinfo {author} {\bibfnamefont {D.}~\bibnamefont
  {Schmid}},\ }\bibfield  {title} {\bibinfo {title} {Time-resolving experiments
  on dicke superfluorescence of $\mathrm{O}_{2}^{}{}_{}{}^{\ensuremath{-}}$
  centers in kcl. two-color superfluorescence},\ }\href
  {https://doi.org/10.1103/PhysRevA.29.2709} {\bibfield  {journal} {\bibinfo
  {journal} {Phys. Rev. A}\ }\textbf {\bibinfo {volume} {29}},\ \bibinfo
  {pages} {2709} (\bibinfo {year} {1984})}\BibitemShut {NoStop}%
\bibitem [{\citenamefont {Loudon}(2000)}]{loudon2000quantum}%
  \BibitemOpen
  \bibfield  {author} {\bibinfo {author} {\bibfnamefont {R.}~\bibnamefont
  {Loudon}},\ }\href@noop {} {\emph {\bibinfo {title} {The quantum theory of
  light}}},\ \bibinfo {edition} {3rd}\ ed.\ (\bibinfo  {publisher} {Oxford
  University Press},\ \bibinfo {year} {2000})\BibitemShut {NoStop}%
\end{thebibliography}%


\begin{thebibliography}{3}%
\makeatletter
\providecommand \@ifxundefined [1]{%
 \@ifx{#1\undefined}
}%
\providecommand \@ifnum [1]{%
 \ifnum #1\expandafter \@firstoftwo
 \else \expandafter \@secondoftwo
 \fi
}%
\providecommand \@ifx [1]{%
 \ifx #1\expandafter \@firstoftwo
 \else \expandafter \@secondoftwo
 \fi
}%
\providecommand \natexlab [1]{#1}%
\providecommand \enquote  [1]{``#1''}%
\providecommand \bibnamefont  [1]{#1}%
\providecommand \bibfnamefont [1]{#1}%
\providecommand \citenamefont [1]{#1}%
\providecommand \href@noop [0]{\@secondoftwo}%
\providecommand \href [0]{\begingroup \@sanitize@url \@href}%
\providecommand \@href[1]{\@@startlink{#1}\@@href}%
\providecommand \@@href[1]{\endgroup#1\@@endlink}%
\providecommand \@sanitize@url [0]{\catcode `\\12\catcode `\$12\catcode
  `\&12\catcode `\#12\catcode `\^12\catcode `\_12\catcode `\%12\relax}%
\providecommand \@@startlink[1]{}%
\providecommand \@@endlink[0]{}%
\providecommand \url  [0]{\begingroup\@sanitize@url \@url }%
\providecommand \@url [1]{\endgroup\@href {#1}{\urlprefix }}%
\providecommand \urlprefix  [0]{URL }%
\providecommand \Eprint [0]{\href }%
\providecommand \doibase [0]{https://doi.org/}%
\providecommand \selectlanguage [0]{\@gobble}%
\providecommand \bibinfo  [0]{\@secondoftwo}%
\providecommand \bibfield  [0]{\@secondoftwo}%
\providecommand \translation [1]{[#1]}%
\providecommand \BibitemOpen [0]{}%
\providecommand \bibitemStop [0]{}%
\providecommand \bibitemNoStop [0]{.\EOS\space}%
\providecommand \EOS [0]{\spacefactor3000\relax}%
\providecommand \BibitemShut  [1]{\csname bibitem#1\endcsname}%
\let\auto@bib@innerbib\@empty
\bibitem [{\citenamefont {Dicke}(1954)}]{dicke1954coherence}%
  \BibitemOpen
  \bibfield  {author} {\bibinfo {author} {\bibfnamefont {R.~H.}\ \bibnamefont
  {Dicke}},\ }\bibfield  {title} {\bibinfo {title} {Coherence in spontaneous
  radiation processes},\ }\href
  {https://journals.aps.org/pr/abstract/10.1103/PhysRev.93.99} {\bibfield
  {journal} {\bibinfo  {journal} {Phys. Rev.}\ }\textbf {\bibinfo {volume}
  {93}},\ \bibinfo {pages} {99} (\bibinfo {year} {1954})}\BibitemShut {NoStop}%
\bibitem [{\citenamefont {Gross}\ and\ \citenamefont
  {Haroche}(1982)}]{gross1982superradiance}%
  \BibitemOpen
  \bibfield  {author} {\bibinfo {author} {\bibfnamefont {M.}~\bibnamefont
  {Gross}}\ and\ \bibinfo {author} {\bibfnamefont {S.}~\bibnamefont
  {Haroche}},\ }\bibfield  {title} {\bibinfo {title} {Superradiance: An essay
  on the theory of collective spontaneous emission},\ }\href
  {https://www.sciencedirect.com/science/article/pii/0370157382901028}
  {\bibfield  {journal} {\bibinfo  {journal} {Phys. Rep.}\ }\textbf {\bibinfo
  {volume} {93}},\ \bibinfo {pages} {301} (\bibinfo {year} {1982})}\BibitemShut
  {NoStop}%
\bibitem [{\citenamefont {Carmichael}(1999)}]{carmichael1999statistical}%
  \BibitemOpen
  \bibfield  {author} {\bibinfo {author} {\bibfnamefont {H.~J.}\ \bibnamefont
  {Carmichael}},\ }\href
  {https://link.springer.com/book/10.1007/978-3-662-03875-8} {\emph {\bibinfo
  {title} {Statistical methods in quantum optics 1: master equations and
  Fokker-Planck equations}}},\ Vol.~\bibinfo {volume} {1}\ (\bibinfo
  {publisher} {Springer Science \& Business Media},\ \bibinfo {year}
  {1999})\BibitemShut {NoStop}%
\end{thebibliography}%

\end{document}


\title{Supplemental Material for\\Emergence of second-order coherence in superfluoresence}

\author{Constanze Bach}
\author{Felix Tebbenjohanns}
\author{Christian Liedl}
\author{Philipp Schneeweiss}
\author{Arno Rauschenbeutel}

\affiliation{Department of Physics, Humboldt-Universität zu Berlin, 10099 Berlin, Germany}

\maketitle

\section{Second-order coherence function of the initial product state}
As described in the main text, we initialize our atomic ensemble at $t=0$ close to the product state given by Eq.~(1),
\begin{equation}
    \ket{\psi_0} = \bigotimes_{k=1}^N \left[\cos \left(  \frac{A}{2}\right) \ket {g_k} - i  \sin \left(  \frac{A}{2}\right) \ket {e_k}\right],
    \label{eq:si_product_state}
\end{equation}
where $\ket{g_k}$ and $\ket{e_k}$ are the ground and excited state of the $\nth{k}$ atom, and $A$ is the Rabi pulse area.
Here, we will compute the first- and second-order correlations of the symmetric lowering operator $\hat S  = \sum_{k=1}^N \spindown{k}$, where $\spindown{k} = \ket{g_k}\bra{e_k}$ is the spin-lowering operator of the $\nth{k}$ atom. 
Since the detected mode $\hat E$ is proportional to $\hat S$, the correlations will hold for $\hat E$ up to a prefactor.
At first, we define $d$ as the imaginary part of the dipole moment, 
\begin{equation}
    d := \imu \braket{\spinup{k} - \spindown{k}} = \imu \bra{\psi_0} \spinup{k} - \spindown{k} \ket{\psi_0} = -2\cos \left(  \frac{A}{2}\right)\sin \left(  \frac{A}{2}\right).
\end{equation}
Note that here, all atoms have the same dipole moment $\braket{\spindown{k}} = \imu d/2$ and therefore we can omit an index $k$.
Next, we define the probability $p$ that one particular atom is excited,
\begin{equation}
    p := \braket{\spinup{k} \spindown{k}} = \sin \left(  \frac{A}{2}\right)^2.
\end{equation}
Since $\ket{\psi_0}$ is a product state, we make use of $\braket{\hat A \hat B} = \braket{\hat A}\braket{\hat B}$ whenever $\hat A$ and $\hat B$ act on different atoms.
With this and $\spindown{k}^2=0$ we find
\begin{subequations}
\begin{align}
    \braket{\hat S} &= \sum_{k=1}^N \braket{\spindown{k}} = \imu N\frac{d}{2} , \\
    \begin{split}
        \braket{\hat S^\dagger \hat S} &= \sum_{k,l=1}^N\braket{\spinup{k}\spindown{l}} = \sum_{k,l=1}^N\braket{\spinup{k}}\braket{\spindown{l}} + \sum_{k=1}^N \left(\braket{\spinup{k}\spindown{k}} - |\braket{\spindown{k}}|^2\right) = |\braket{\hat S}|^2 + N \left(p - \frac{d^2}{4}\right) \\
        & = N \left[p + (N-1)  \frac{d^2}{4}\right] \xrightarrow{N\gg 1} N \left(p + N  \frac{d^2}{4}\right)
    \end{split} \\
    \begin{split}
        \braket{\hat S^\dagger \hat S^\dagger \hat S \hat S} &= \sum_{k,l,m,n=1}^N\braket{\spinup{k}\spinup{l}\spindown{m}\spindown{n}} = \sum_{k,l,m,n=1}^N
         \begin{cases} \frac{d^4}{16}, & \text{$k,l,m,n$ all different} \\
         p \frac{d^2}{4}, & \text{either $k$ or $l$ equals either $m$ or $n$} \\
         p^2, & \text{$k=m\neq l=n$ or $k=n\neq l=m$} \\
         0, & \text{else}
        \end{cases}
        \\
        &=  N(N-1)\left[(N-2)(N-3)\frac{d^4}{16} + 4(N-2) p \frac{d^2}{4} +  2p^2 \right]\\
        &\xrightarrow{N \gg 1} N^2\left(N^2\frac{d^4}{16} + 4p N  \frac{d^2}{4} +  2p^2 \right)
    \end{split}.
\end{align}
\end{subequations}
Here, the final expressions are valid in the large-ensemble limit of $N\gg1$. In that limit, we can express the second-order correlation function as
\begin{equation}
    \braket{\hat S^\dagger \hat S^\dagger \hat S \hat S}
    = \braket{\hat S^\dagger \hat S}^2 \left(2 - \frac{ N^4 d^4}{16\braket{\hat S^\dagger \hat S}^2}\right) = \braket{\hat S^\dagger \hat S}^2 \left(2 - \left(1+\frac{4p}{Nd^2} \right)^{-2}\right).
\end{equation}
With the expressions for $p$ and $d$, we thus find for the normalized second-order coherence function
\begin{equation}
    g^{(2)}(0,0) =\frac{\braket{\hat S^\dagger \hat S^\dagger \hat S \hat S}}{\braket{\hat S^\dagger \hat S}^2}= 2 - \left(1+\frac{1}{N \cos^2(A/2)} \right)^{-2},
\end{equation}
which is Eq.~(3) of the main text.
Let us analyze this expression as a function of $A$.
First note that it has a period of $2\pi$ in $A$. Next, on the interval $(0,2\pi)$ the function has one maximum at $A=\pi$ with $g^{(2)}(0,0) = 2$. Furthermore, for $A\neq \pi$, the function approaches $1$ as $N$ becomes large, i.e. $g^{(2)}(0,0) \xrightarrow{N\to\infty} 1$. Thus, for a finite $N\gg 1$, the function is similar to a Lorentzian with a narrow peak at $A=\pi$, whose full-width at half-maximum (FWHM) $w$ can be computed as follows. 
We set $A=\pi \pm w/2$ and have in close proximity to the maximum (i.e. for $w\ll\pi$), $g^{(2)}(t,t)\approx2-[1+16/(N w^2)]^{-2}$. From this, we find the FWHM as 
\begin{equation}
    w = \sqrt{\frac{16}{\sqrt{2}-1} \frac{1}{N}} \approx 0.9892\times \frac{2\pi}{\sqrt{N}}.
\end{equation}

\section{Two-time second-order correlation function in the Dicke model} \label{sec:Dicke}

\begin{figure}
    \centering    \includegraphics[width=0.7\columnwidth]{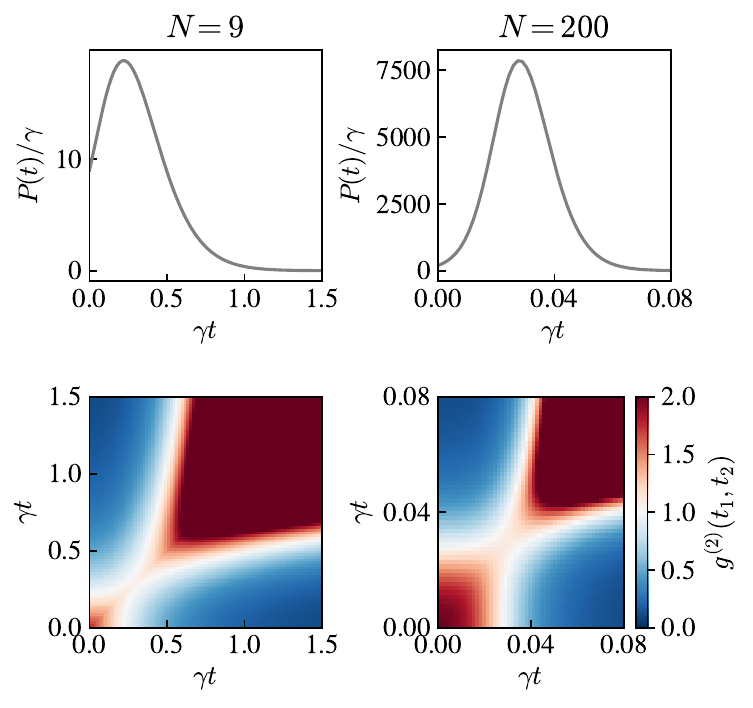}
    \caption{Symmetric Dicke model predictions for $N=9$ (left) and $N=200$ (right) identical two-level atoms with excited state lifetime $\tau=1/\gamma$. Top row: normalized photon flux $P(t)/\gamma = \braket{\hat S^\dagger(t) \hat S(t)}$ as a function of time. Bottom row: color plot of $g^{(2)}(t_1, t_2)$. Red (blue) colors indicate $g^{(2)}>1$ ($g^{(2)}<1$).}
    \label{fig:sub}
\end{figure}

In this section, we consider the original Dicke model~\cite{dicke1954coherence,gross1982superradiance}. 
There, a very dense ensemble of $N$ two-level atoms with excited state lifetime $\tau=1/\gamma$ is initialized in the fully inverted state $\ket{N} = \ket{e\cdots e}$. Upon decay, this ensemble always stays in its symmetric Dicke states $\ket{k}$, which can be written as~\cite{gross1982superradiance}
\begin{equation}
    \ket{k} = \sqrt{\frac{k!}{N!(N-k)!}} \hat S^{N-k} \ket N,
\end{equation}
where $k=0,1,\cdots,N$ counts the number of excitations in the ensemble and $\hat S =\sum_{k=1}^N \spindown{k}$ is the symmetric lowering operator.
Our goal is to derive an expression for the second-order two-time correlation function $G^{(2)}(t_1,t_2)=\braket{\hat E^\dagger(t_1)\hat E^\dagger(t_2) \hat E(t_2) \hat E(t_1)}$ of the field $\hat E\propto\sqrt{\gamma} \hat S$, radiated by the Dicke ensemble. 

For this, consider the vector of operators $\vec{\hat x_0}$, whose elements are the projection operators $(\vec{\hat x_0})_k =\ket{k}\bra{k}$. Note that the expected value of $\vec{\hat x_0}$ are the diagonal entries of the density matrix with $\rho_{kk} = \bra{k}\hat\rho\ket{k} = (\braket{\vec{\hat x_0}})_k$. The equation of motion of these diagonal entries reads~\cite{gross1982superradiance}
\begin{equation}\label{eq:si_deq_diagonals}
\frac{\diff}{\diff t} \braket{\vec{\hat x_0}}= \gamma A_0 \braket{\vec{\hat x_0}}
\end{equation}
with the $(N+1)^2$ - matrix $A_0$, whose components are given by ($m,n=0,\dots,N$)
\begin{equation}
    \left(A_0\right)_{nm} = -\delta_{nm}s_n^2 + \delta_{n+1,m}s_{n+1}^2.
\end{equation}
Here, $s_k=\sqrt{k(N+1-k)}$ are defined through $\hat S \ket{k} = s_k \ket{k-1}$ and $\hat S^\dagger \ket{k}  = s_{k+1} \ket{k+1}$.
The time-dependent solution of Eq.~\eqref{eq:si_deq_diagonals} reads  $\rho_{kk}(t) = \braket{\vec{\hat x_0}(t)} = \expu^{A_0 \gamma t} \rho_{kk}(0)$ with the initial value $\rho_{kk}(0) = \delta_{k,N}$.
Note that a numerical solution to this matrix exponential can be found in a reasonable time, even when $N$ exceeds $100$.
The off-diagonal elements are all zero, $\rho_{k\neq n}(t)=0$.
The time-dependent radiated photon flux $P(t)$ is then found by using $\hat S^\dagger \hat S = \sum_{k=0}^N s_k^2 (\vec{\hat x_0})_k$ as
\begin{equation}
    P(t) = \gamma \braket{\hat S^\dagger(t) \hat S(t)} = \gamma \tr \left[\hat\rho(t) \hat S^\dagger \hat S\right] = \gamma \sum_{k,n=0}^N  s_k^2 \bra{n} \hat \rho(t) (\vec{\hat x_0})_k \ket{n} =\gamma  \sum_{k=0}^N s_k^2 \rho_{kk}(t).
\end{equation}
We now consider the vector of two-time correlators $\braket{\hat S^\dagger(t_1)  \vec{\hat x_0(t_2)} \hat S(t_1)}$ with $t_2>t_1$. 
Employing the quantum regression theorem~\cite{carmichael1999statistical}, we find an equation of motion for this vector of correlators from Eq.~\eqref{eq:si_deq_diagonals} as
\begin{equation}
    \frac{\diff}{\diff t_2}  \braket{\hat S^\dagger(t_1)  \vec{\hat x_0(t_2)} \hat S(t_1)} = \gamma A_0  \braket{\hat S^\dagger(t_1)  \vec{\hat x_0(t_2)} \hat S(t_1)}.
\end{equation}
The solution to this differential equation reads $ \braket{\hat S^\dagger(t_1)  \vec{\hat x_0(t_2)} \hat S(t_1)} = \expu^{A_0 \gamma (t_2-t_1)} \braket{\hat S^\dagger(t_1)  \vec{\hat x_0(t_1)} \hat S(t_1)}$.
From this, we finally find the two-time second-order correlation function as 
\begin{equation}\label{eq:si_G2_twotime}
\begin{split}
& G^{(2)}(t_1,t_2) = \braket{\hat S^\dagger(t_1)\hat S^\dagger(t_2) \hat S(t_2) \hat S(t_1)} =  \sum_{k=0}^N s_k^2 \left( \braket{\hat S^\dagger(t_1)  \vec{\hat x_0(t_2)} \hat S(t_1)} \right)_k \\
&=  \sum_{k,n=0}^N s_k^2 \left( \expu^{A_0 \gamma (t_2-t_1)} \right)_{k,n} \braket{\hat S^\dagger \ket{n}\bra{n} \hat S}_{t_1} 
=  \sum_{k,n=0}^N s_k^2 s_{n+1}^2 \left( \expu^{A_0 \gamma (t_2-t_1)} \right)_{k,n}  \left(\vec{\rho}_0(t_1)\right)_{n+1}.
\end{split}
\end{equation}
For $t_2<t_1$, we use the symmetry property $G^{(2)}(t_1,t_2) = G^{(2)}(t_2,t_1)$.
The normalized second-order coherence function is then found by
\begin{equation}
    g^{(2)}(t_1,t_2) = \frac{G^{(2)}(t_1,t_2)}{P(t_1)P(t_2)}.
\end{equation}
We show this quantity as a color plot for $N=9$ and $N=200$ in Fig.~\ref{fig:sub}.

\bibliography{bib}